\documentclass[aps,prb,twocolumn,superscriptaddress]{revtex4-1}

\usepackage{amsmath}
\usepackage{amssymb}
\usepackage{graphicx}
\usepackage[utf8]{inputenc}
\usepackage{braket}
\usepackage[%
  colorlinks=true,
  urlcolor=blue,
  linkcolor=blue,
  citecolor=blue
]{hyperref}
\usepackage{color}
\usepackage[table]{xcolor}

\begin{document}

\title{Role of impurity clusters for the current-driven motion of magnetic Skyrmions}

\author{M. Stier}
\affiliation{I. Institut für Theoretische Physik, Universität Hamburg, Jungiusstraße 9, 20355 Hamburg, Germany}
\author{R. Strobel}
\affiliation{I. Institut für Theoretische Physik, Universität Hamburg, Jungiusstraße 9, 20355 Hamburg, Germany}
\author{W. Häusler}
\affiliation{Institut für Physik, Universität Augsburg, Universitätsstraße 1, 86159 Augsburg, Germany}

\author{M. Thorwart}
\affiliation{I. Institut für Theoretische Physik, Universität Hamburg, Jungiusstraße 9, 20355 Hamburg, Germany}

\date{\today}

\begin{abstract}
We study how impurities influence the current-induced dynamics of magnetic Skyrmions moving in a racetrack geometry. For this, we  solve numerically the generalized Landau-Lifshitz-Gilbert equation extended by the current-induced spin transfer torque. In particular, we investigate two classes of impurities, non-conducting and magnetic impurities. The former are magnetically rigid objects and yield to an inhomogeneous current density over the racetrack which we determine separately by solving the fundamental electrostatic equations. In contrast, magnetic impurities
leave the applied current density homogeneous throughout the stripe. Depending on
parameters, we observe four different scenarios of Skyrmion
motions in the presence of disorder, the Skyrmion decay, the pinning, the creation of additional
Skyrmions, and ordinary Skyrmion passage. We calculate and discuss phase diagrams in dependence of the impurity concentration and
radii of the impurities.
\end{abstract}


\maketitle 
\section{Introduction}
Their topological protection renders magnetic Skyrmions interesting candidates for reliable processing of information by racetrack geometries. Skyrmions are vortex-like spin textures that can
arise in non-centrosymmetric magnetic compounds. They are rather stable, extremely small in size and can be moved by low densities of spin-polarized electronic currents.
For example, the interplay of Heisenberg exchange interaction, antisymmetric Dzyaloshinskii-Moriya interaction (DMI), and an external Zeeman field may
favor the formation of magnetic Skyrmions. They have been predicted\cite{bogdanov1, bogdanov1, Bogdanov1995} at the beginning of the 1990s and have been realized  experimentally in magnetic layers with a strong spin-orbit interaction\cite{ pfleidererrosch2, yu1, heinze}. Skyrmions can be moved under the action of a spin transfer torque. This is induced by flowing electrons in an externally applied spin-polarized current at current densities of the order $10^5$ $\text{A}/\text{m}^2$\cite{nagaosa1, Lu2014}. Particularly this latter feature makes Skyrmions very promising candidates for future spintronic applications, especially when using racetrack memories consisting of thin nanowires.

Topological protection of the Skyrmions is characterized by a nonzero, integer valued topological charge $Q$, also called Skyrmion number\cite{romming}. As an idealized concept originating from continuum theory it is
considered to protect Skyrmions against decay, even in the
presence of imperfections in the host material as a result of
the fabrication process. In reality, the argument of
topological stability has to be replaced by a finite energy
barrier when the objects exist on a lattice\cite{garanin18,
rosch19,fuchsbacharb}. Moreover, real experimental situations need to be described as open systems, especially when externally applied spin-polarized currents are applied. Then, dissipation induced by external fluctuations can play a central role, for instance, also for the creation of Skyrmions. It has been shown \cite{stier} that Skyrmions can be generated by an external current such that first a charge neutral Skyrmion-Antiskyrmion pair is created. Since the Antiskyrmion is unstable under the action of dissipation, it decays over time and only the Skyrmion partner survives. By this nonequilibrium process, topological charge is created as a result of dissipation.

Since topological protection is an idealized concept, a valid question is about the role of disorder impurities for the current-induced Skyrmion dynamics in real finite-size lattice systems. This question appears also in the context of the Skyrmion Hall effect whose interplay with disorder has not been studied so far. Current-driven Skyrmions experience the Skyrmion Hall effect, in which they develop a motion perpendicular to the direction of the applied current, just like charged particles in the standard Hall effect\cite{Litzius2017, jiangshe}. The presence of disorder could in principle influence the Skyrmion Hall effect, reduce or even enhance it. 

Current-induced Skyrmion motion in the presence of disorder has been studied experimentally in a few material configurations. In a racetrack designed in a layered Pt/FM/Au/FM/Pt-system \cite{realhrabec}, individual Skyrmions covered a different spatial distance although they were all driven by the same current pulses. They got stuck in the racetrack eventually. The authors attributed this behavior to the presence of impurities. In a comparable experiment \cite{realjuge} with a racetrack on the basis of a Ta/Pt/Co/MgO/Ta film, the spatial dimensions of the Skyrmions were of the order of  $\sim 100$nm. While some Skyrmions traveled several $100$nm long distances under the action of the same pulses, some got destroyed and in some cases, new Skyrmions appeared. Also here, impurities were hold responsible for this behavior. 

In a further study of a racetrack on the basis of a Pt/Co/Ta layered system, Skyrmions were also driven by current pulses \cite{realwoo}. In this set-up, some Skyrmions got pinned and could not be moved further. Also Skyrmion decay was observed. In addition, Skyrmions in different regions of the racetrack moved with different velocities under the action of identical current pulses. In comparison to a Pt/CoFeB/MgO racetrack, where the effect of impurities were expected to be less detrimental, a more regular motion of the Skyrmions could be detected, together with much less frequent pinning. 

Motivated by this great variety of different behaviors, we study in this work the role of impurities for the current-induced dynamics of magnetic Skyrmions. For this we solve numerically the generalized Landau-Lifshitz-Gilbert equation extended by the current-induced spin transfer torque. We study two classes of impurities below, non-conducting and magnetic impurities. The former are rigid objects and yield to an inhomogeneous current density which acts throughout the racetrack. It has to be determined separately by solving the fundamental electrostatic equations. In contrast, magnetic impurities allow for a homogeneous applied current density which is constant throughout the stripe. Depending on the parameter configurations, we observe four different types of Skyrmion motions in the presence of disorder, the Skyrmion decay, the pinning, the creation of additional Skyrmions, and the ordinary Skyrmion passing. Phase diagrams in dependence of the impurity concentration and the radius of impurity clusters are calculated and discussed in this work. 

\section{Model and magnetization dynamics}

We study the dynamics of a field of unit magnetization vectors ${\bf n} ({\bf r},t)=\bf{M} ({\bf r},t)/| \bf{M} ({\bf r},t)|$ on a two-dimensional square lattice in the $xy$-plane with lattice constant $a$. A Skyrmion is stabilized by the Hamiltonian
\begin{equation}
	\begin{split}
	{\cal H} &
	= {\cal H}_{\text{ex}} + {\cal H}_{\text{DM}} + {\cal H}_{\text{Zeeman}} + {\cal H}_{\text{aniso}}\\\
	&=-J \sum_{{\bf r}} {\bf n}  ({\bf r}) \cdot \big[ {\bf n}  ({\bf r} + a\cdot {\bf e}_{x}) + {\bf n} ({\bf r} + a\cdot {\bf e}_{y}) \big] \\\
	&\ \ \ \ - D \sum_{{\bf r}} \left\{ \big[ {\bf n} ({\bf r}) \times {\bf n}  ({\bf r} + a \cdot {\bf e}_{x}) \big] \cdot {\bf e}_{x} \right.  \\\
	&\ \ \ \ \ \ \left.  + \big[ {\bf n}  ({\bf r}) \times {\bf n} ({\bf r} + a \cdot {\bf e}_{y}) \big] \cdot{\bf e}_{y} \right\} \\\  
	&\ \ \ \  -\gamma \hbar {\bf B}  ({\bf r}) \cdot \sum_{{\bf r}} {\bf n}  ({\bf r}) - \sum_{\bf r} K({\bf r}) n^{2}_{z} ({\bf r}).     \end{split}
\label{eq:hamiltonOperator}
\end{equation}
As usual, the effective magnetic field is defined via the relation ${\cal H} = -\gamma \hbar \sum_{\bf r} {\bf B}_{\text{eff}} ({\bf r}) \cdot {\bf n} ({\bf r})$, such that ${\bf B}_{\text{eff}} ({\bf r}) =-\frac{1}{ {\gamma \hbar}}\frac{\delta {\cal H}}{\delta {\bf n} ({\bf r})}$, with the gyromagnetic ratio $\gamma$.  The last term describing the magnetic anisotropy $K({\bf r})$ is nonzero only inside of magnetic impurity clusters.

For concreteness, we consider throughout this work a racetrack geometry with $512 \times 128$ lattice points and $a = 0.5$ nm. We choose the parameters  $J=1 \text{meV}$, $D=0.18 \text{meV}$ und ${\gamma\hbar\bf B} = -0.025 \text{meV} {\bf e}_{z}$. This choice mimics a surface of MnSi, which is a chiral magnet with a DMI, which stabilizes Bloch Skyrmions \cite{nagaosa1}.

For the racetrack geometry with borders in both directions, we use periodic boundary conditions in the longitudinal $x$-direction, i.e., the spins ${\bf n}_{1,j}(t)$ and ${\bf n}_{512,j}(t)$ are neighbors. For those lattice points which do not have direct neighbors, e.g., due to lateral confinement or impurities, we use ferromagnetic boundary conditions, i.e., the missing neighboring magnetic moment is replaced by $-{\bf e}_{z} = (0,0,-1)$. This boundary condition fixes the boundary magnetization to the direction of the external magnetic field.

The dynamics of the magnetization field is described by the generalized Landau-Lifshitz-Gilbert equation (LLG)  
\begin{equation}
\begin{split}
	\frac{\partial {\bf n} ({\bf r})}{\partial t} = 
	& \bigg\{ -\gamma {\bf n} ({\bf r})\times {\bf B}_{\text{eff}} ({\bf r})-\alpha \gamma {\bf n} ({\bf r}) \times \big[ {\bf n} ({\bf r})\times{\bf B}_{\text{eff}} ({\bf r}) \big] \\\
	& + (\alpha - \beta) {\bf n} ({\bf r}) \times \big[ \left({\bf v}_{\text{s}} \cdot \nabla\right) {\bf n} ({\bf r}) \big] \\\
	&       + (\alpha \beta +1) \left( {\bf v}_{\text{s}} \cdot \nabla\right) {\bf n} ({\bf r})\bigg\} \frac{1}{1+\alpha^{2}} \, ,
	\end{split}     
\label{eq:erweitertellg}
\end{equation}
with the Gilbert damping parameter $\alpha$, and the nonadiabaticity parameter $\beta$.
Throughout this work, we choose  $\alpha = 0.1$ which is in a realistic range of the values $\alpha = 0.02$ to $0.4$ for CoPt bilayers  \cite{barati}. Moreover, we choose two different values for the nonadiabaticity parameter  $\beta = 0.05$ and $\beta = 0.1$. By this, we can compare the situations with ($\beta / \alpha = 0.5$) and without ($\beta / \alpha = 1$) the Skyrmion Hall effect.  Those values are motivated by the case of Permalloy, for which we have $\beta = 0.02$ to $0.12$ \cite{sekiguchi,roessler}.

The externally applied charge current density ${\bf j}_{\text{c}}$
(elementary charge $e>0$) determines the velocity
\begin{equation}
{\bf v}_{\text{s}} ({\bf r}) = \frac{pa^{3}}{2e} {\bf j}_{\text{c}} ({\bf r})
\label{eq:spinpolarisierterStrom}
\end{equation}
of effective spin carriers, entering Eq.\ (\ref{eq:erweitertellg}). Here, $p$ denotes the spin polarization. The LLG contains the adiabatic as well as the nonadiabatic contribution to the current-induced spin transfer torque. The last term induces a drift of the magnetic texture in the direction of the current, while the third term leads to a displacement perpendicular to the current density. This term is responsible for the Skyrmion Hall effect \cite{stier} and vanishes for the special case $\alpha = \beta$.

As a starting configuration ${\bf n} (x,y,t=0)$, we choose one Bloch Skyrmion with the topological charge $N_{\text{sk}} = 1$, placed at $x_{\text{sk}} = 63.8a$ und $y_{\text{sk}} = 67.4a$ close to the left border of the racetrack stripe. It is dynamically relaxed without any current and without any impurity clusters to a radius of $R_{\text{sk}} \approx 7.7a$, cf.\ Eq.\ (\ref{eq:skradius}). The initial velocity at the left border is fixed to
$v_{\text{s}} = -0.05$ in units of $0.5\text{nm}/\text{ps}$, which corresponds to a physical current density of  $j_{\text{c}} = 6.4 \times 10^{10}\frac{\text{A}}{\text{m}^2}$ (choosing ideal polarization $p=1$). The discretized extended LLG is integrated over time by a standard fourth-order Runge-Kutta scheme in time steps of $\Delta t = 0.01 \text{ps}$. It terminates usually at  $t_{\text{end}} = 10000 \text{ps}$ and has an error-controlled adaptive time discretization.

\begin{figure}[t!]
\includegraphics[width=0.99\linewidth]{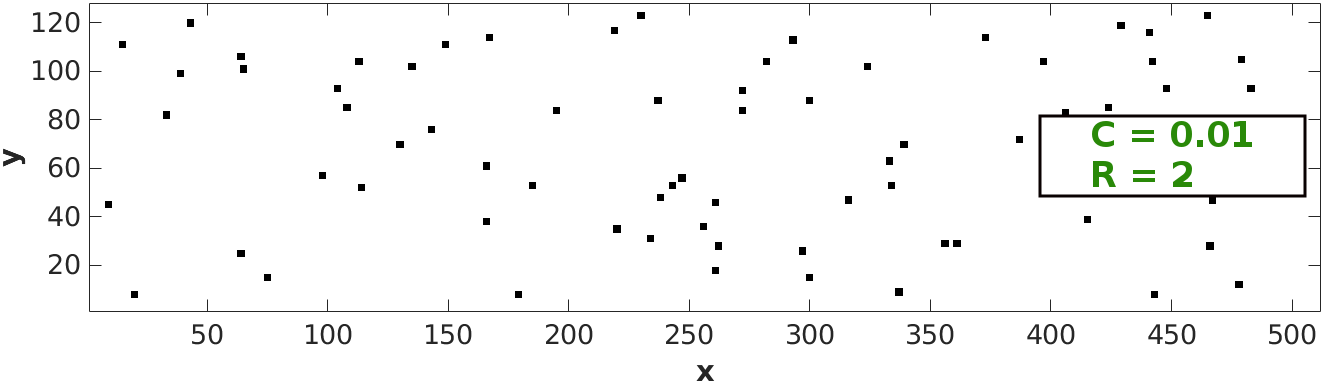}\\
\includegraphics[width=0.99\linewidth]{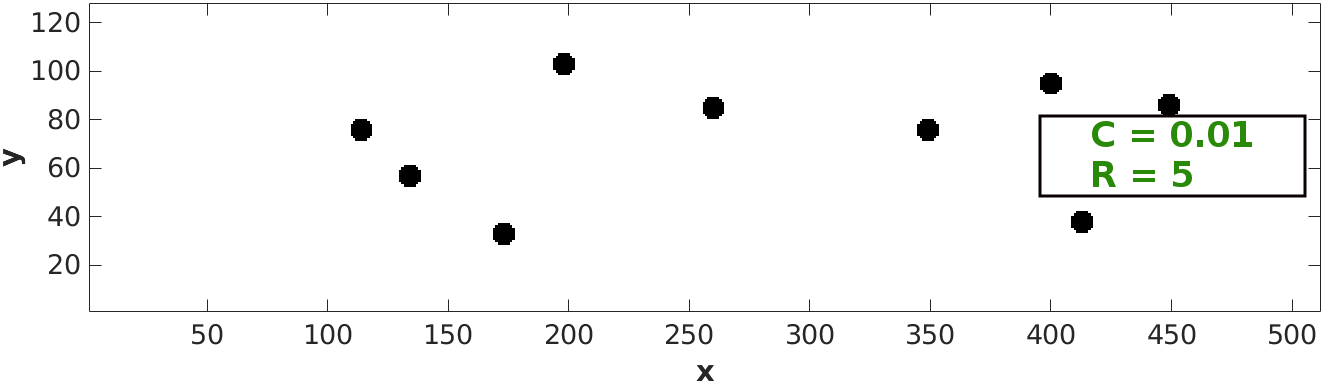}\\ 
\includegraphics[width=0.99\linewidth]{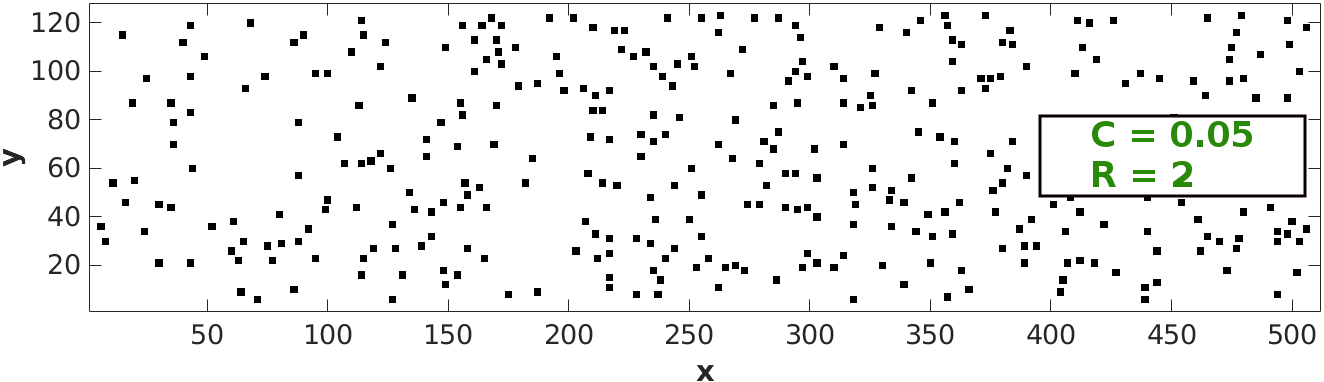}\\ 
\includegraphics[width=0.99\linewidth]{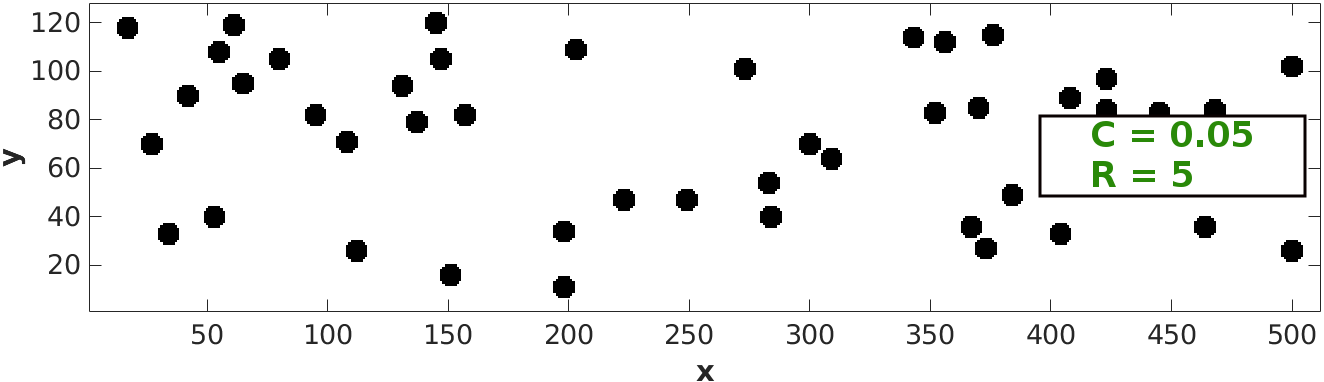}
\caption{\label{fig1} 
Examples of racetracks of a $512 \times 128$-lattice, containing impurity clusters. The impurities are shown in black and configurations are characterized by the cluster radius $R$ and the total concentration $C$. }
\end{figure}

With the resulting solution $ {\bf n} ({\bf r},t)$, we calculate the topological charge density
\begin{equation}
	q(x,y,t) = q({\bf r},t) = - \frac{1}{4\pi} {\bf n} ({\bf r},t) \cdot \bigg[ \frac{\partial {\bf n} ({\bf r},t)}{\partial x} \times \frac{\partial {\bf n} ({\bf r},t)}{\partial y} \bigg]\, ,
\label{eq:topLadungsdichte}
\end{equation}
where the minus sign appears due to the choice of the ferromagnetic background ${\bf n}_{\text{hom}}=-{\bf e}_{z}$ in this work. This results in a positive charge for a Skyrmion and a negative one for an Antiskyrmion. In addition, we calculate the topological charge 
\begin{equation}
N_{\text{sk}}(t)= \int d^{2}r\;\;q(x,y,t) \, ,
\label{eq:topLadung}
\end{equation}
or, the winding number of the Skyrmion. We identify the time dependent position of the Skyrmion via the ``center of mass'' of the topological charge density, which is determined by the coordinates
\begin{eqnarray}
x_{\text{sk}}(t) &=& \sum_{i,j} \frac{a\;i\; q(i,j,t)}{N_{\text{sk}}(t)}      \nonumber \\
y_{\text{sk}}(t) &= & \sum_{i,j} \frac{a\;j\; q(i,j,t)}{N_{\text{sk}}(t)} \, , 
\end{eqnarray}
and likewise define its radius according to 
\begin{equation} 
R_{\text{sk}}(t) = \sqrt{\int dx\:dy\ \left\{[x - x_{\text{sk}}(t)]^2 + [y - y_{\text{sk}}(t)]^2\right\}\frac{q(x,y,t)}{N_{\text{sk}}(t)}}\ . 
\label{eq:skradius} 
\end{equation}
For simplicity we only consider spherical clusters of definite radius $R$. Their positions are randomly equally distributed over the racetrack stripe. Examples of configurations are shown Fig.\ \ref{fig1}. Each disorder configuration is characterized by the radius $R$ of a single impurity cluster and the total concentration $C$, which is defined as the sum of all lattice points to which an impurity belongs divided by the total number of lattice sites. This definition implies that a given $C$, and therefore given
number of impurity lattice points, is realized with fewer
impurity clusters at larger radius $R$. Thus, the distance between
the impurity clusters increases when $R$ increases.

\section{Inhomogeneous current density}
\label{inhomo}

In the case that the racetrack arrangement is altered by inserting nonconducting impurities, the spatial distribution of the applied current density distribution is modified and deviates from a homogeneous distribution. Hence, we need to determine explicitly the charge current density distribution for the case of nonconducting impurities (for conducting impurities, also considered below, the current density stays homogeneous and no additional calculation is required).  

For this, we start from the relevant boundary conditions for the current density vector ${\bf j}_{\text{c}}$. At both longitudinal ends of the racetrack, $x = x_{\text{min}}$ and $x = x_{\text{max}}$, we impose an external current density in the $x$-direction according to ${\bf v}_{\text{s}} (x_{\text{min,max}},y) = (v_{\text{s}},0)$ for all  $y \in [y_{\text{min}},y_{\text{max}}]$. We denote this as the entrance current density. At any border of the racetrack or of an impurity, we impose that the current component normal to the boundary vanishes. Overall, current can only enter via the left and leave the racetrack via the right edge. 

The inhomogeneous current density is calculated from the
electric field distribution by assuming a constant
conductivity $\sigma$ and Ohm's law, ${\bf j}_{\text{c}}
({\bf r}) = \sigma {\bf E}({\bf r})$. The electric field,
in turn, is obtained from the electrostatic potential,
${\bf E}({\bf r})=-\nabla \phi ({\bf r})$, for which we solve
the Laplace equation $\Delta \phi ({\bf r}) = 0$ with Neumann
boundary conditions at all surfaces $\tilde{\bf r}$
\begin{equation}
\frac{\partial \phi (\tilde{\bf r})}{\partial {\bf u}(\tilde{\bf r})} = \nabla \phi (\tilde{\bf r}) \cdot {\bf u}(\tilde{\bf r}) = - {\bf E}(\tilde{\bf r}) \cdot {\bf u}(\tilde{\bf r}) =  -\frac{{\bf j}_{\text{c}}(\tilde{\bf r})}{\sigma} \cdot {\bf u}(\tilde{\bf r})
=0\, .
 \label{eq:neumannRB}
\end{equation}
Here, ${\bf u}(\tilde{\bf r})$ denotes the unit outer normal
vector at all boundary surfaces. With the solution $\phi ({\bf
r})$ at hand, we can then calculate the current density
according to ${\bf j}_{\text{c}} ({\bf r}) / \sigma = {\bf
E}({\bf r}) = -\nabla \phi ({\bf r})$.
\begin{figure}[t!]
\includegraphics[width=0.99\linewidth]{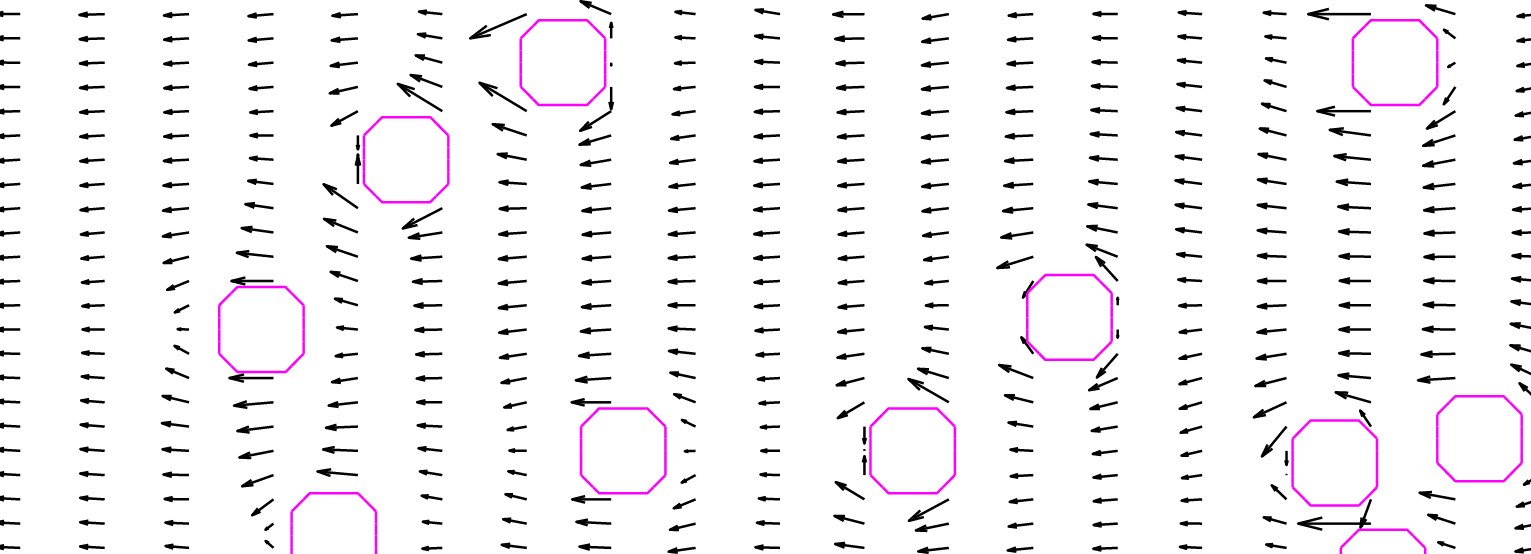}
\caption{\label{fig2} 
Section of the velocity distribution ${\bf v}_{\text{s}} ({\bf r})$, which is proportional to the current density ${\bf j}_{\text{c}} ({\bf r})$,
of a typical racetrack with nonconducting impurities with $C = 0.03$ and $R= 4$. }
\end{figure}

\begin{figure}[t!]
	\includegraphics[width=0.99\linewidth]{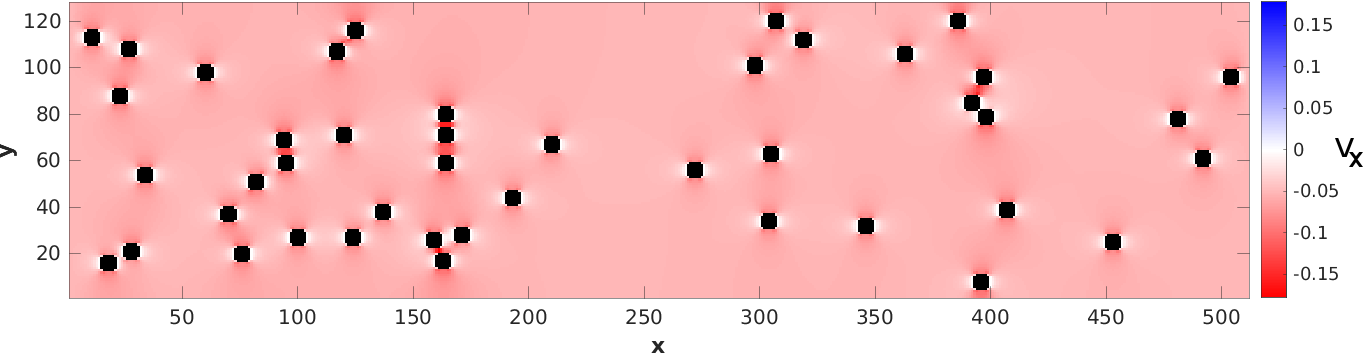}
	\includegraphics[width=0.99\linewidth]{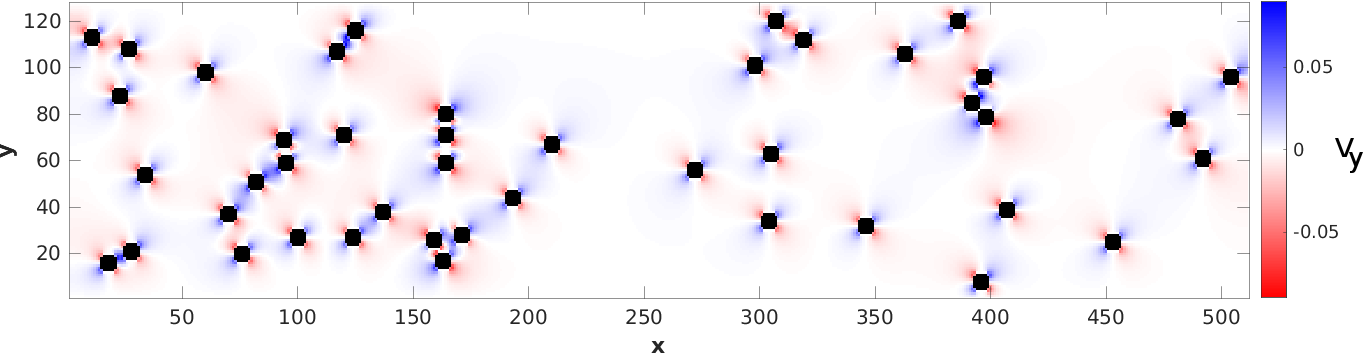}
\caption{\label{fig3} Distribution of the longitudinal (a) and transversal (b) velocity components in the racetrack of Fig.\ \ref{fig2}. }
\end{figure}

\begin{figure}[t!]
	\includegraphics[width=0.9\linewidth]{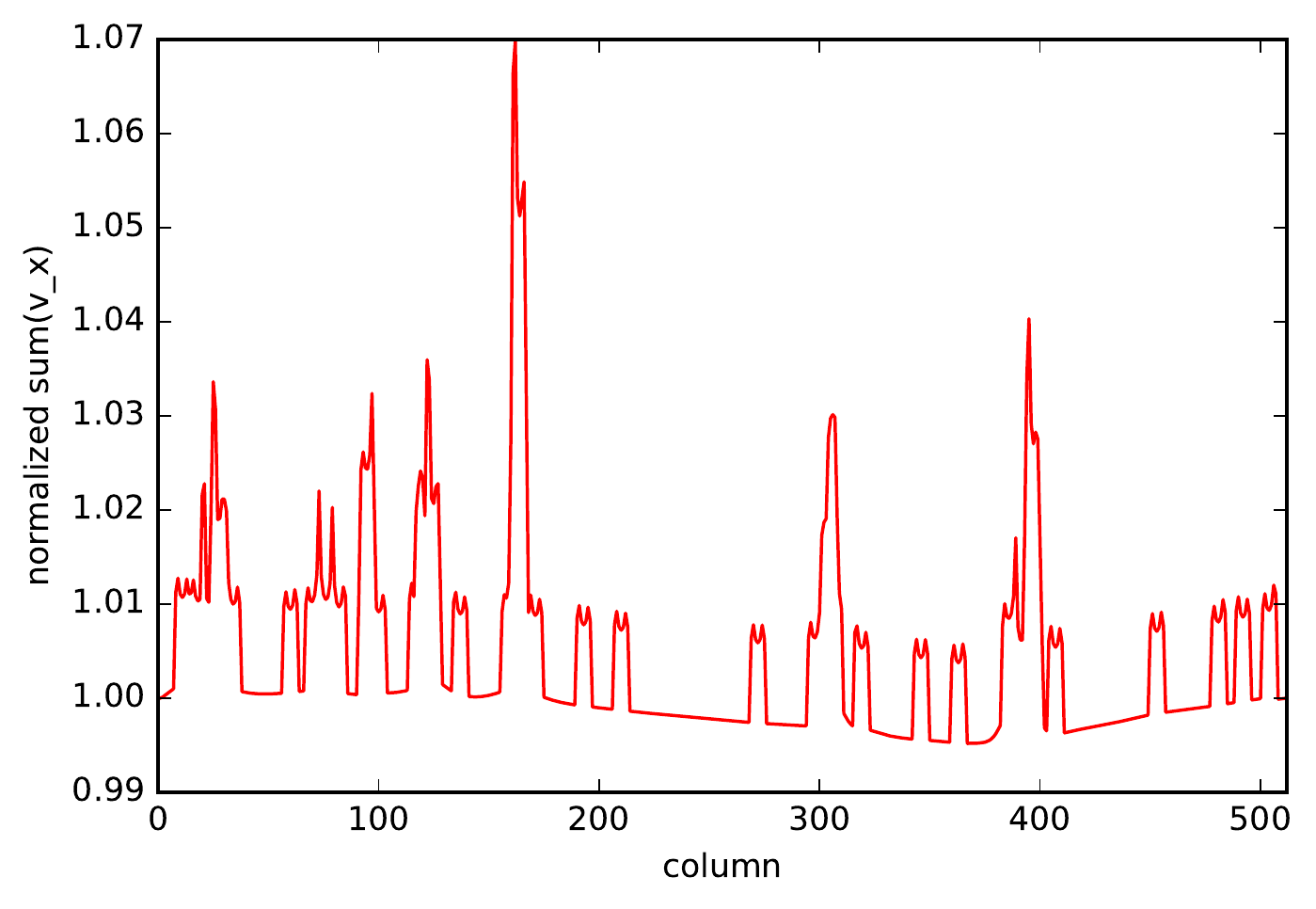}
\caption{\label{fig4} Normalized sum of the charge carriers along the vertical direction (column) of the current density associated to Figs.\ \ref{fig2} and \ref{fig3}.}
\end{figure}

A typical example of a racetrack with nonconducting impurity clusters and an inhomogeneous current density is shown in Fig.\ \ref{fig2}. The distribution of the longitudinal (a) and transversal (b) velocity components of this racetrack is shown in  Fig.\ \ref{fig3}. Due to the use of a continuum theory (electrostatics) in combination with a discrete lattice whose lattice constant is fixed by the given magnetic texture, which cannot be refined arbitrarily in the numerical practice, we encounter the effect that the continuity equation for the charge carriers cannot be fulfilled to full extent. This can be seen by summing the charges over all vertical lattice points (column) for a fixed horizontal coordinate. While we always find that the normalized sum equals 1 at both ends of the racetrack, deviations occur in the interior of the stripe. A typical case is shown in Fig.\ \ref{fig4} which shows the normalized vertical sum of the charges along the horizontal direction. In this case, the deviations occur in the range of a few percent. In general, the deviations grow, the more narrow a channel is. However, they can be significant for small values of $R$ and high concentrations $C$. These pathological cases appear in figures below, but are marked correspondingly.

\section{Classification of the Skyrmion dynamics}

In general, we find four different types of behavior of the Skyrmion which we classify as follows:

{\em Type I: Skyrmion creation}: During the simulation at least a second Skyrmion is created in addition to the initial one.

{\em Type II: Skyrmion decay}: During the simulation, the initial  Skyrmion disappears and no further one is created.

{\em Type III: Skyrmion pinning}: During the simulation, the initial Skyrmion reaches a location where it gets pinned. Neither does this Skyrmion decay nor is a new one created. 

 {\em Type IV: Skyrmion passing}: The initial Skyrmion passes through the racetrack stripe without falling in one of the other three categories. 

We classify the individual configurations according to the following rules: If at a certain time $t$ the charge exceeds $N_{\text{sk}}(t) > 1.5$, the simulation is attributed to type I, Skyrmion creation. When $N_{\text{sk}}(t) < 0.5$, the case is attributed to type II, Skyrmion decay. If both cases occur during one run, the earlier occurrence decides about the classification. 

In all other cases, only one Skyrmion exists throughout the simulation and we trace its $x$-coordinate $x_{\text{sk}}(t)$. When its change over a time span of $1000\text{ps}$ does not exceed a distance of $5a$, we assign it to type III, Skyrmion pinning. Otherwise, it falls into type IV, Skyrmion passing. 

\section{Nonconducting impurity clusters}

\begin{figure}[t!]
	\includegraphics[width=0.99\linewidth]{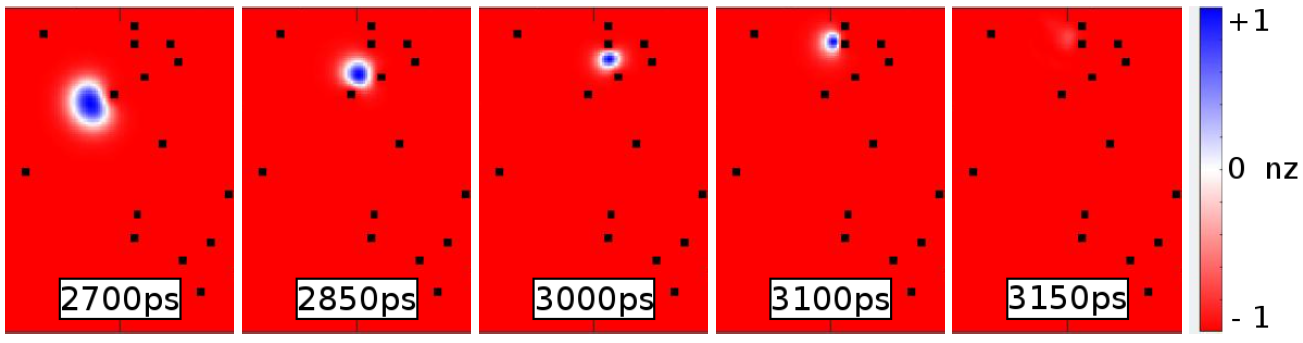}
\caption{\label{fig5} Snapshots of a Skyrmion (blue) moving in a ferromagnetic racetrack (red) with nonmagnetic impurity clusters (black) for $\beta=\alpha$, the initial current density $v_{\text{s}} = -0.05$, the impurity concentration $C = 0.01$ and the impurity radius $R = 2$. At $t = 2700\text{ps}$ and $t = 2850\text{ps}$, it moves around an impurity and tries to do so for a neighboring one as well ($t = 3000\text{ps}$ and $t = 3100\text{ps}$). Yet, it is pressed against it, and, having been compressed already before, it is now further compressed and destroyed at $t = 3150\text{ps}$.}
\end{figure}

In this section, we present the results for nonconducting impurity clusters which induce an inhomogeneous current density determined according to Sec.\ \ref{inhomo}. The impurity clusters form rigid obstacles which cannot be magnetized. On the other hand, the current density can in general cause the Skyrmion to snake around an impurity and to be accelerated on its way. A representative of a Skyrmion trace is shown in Fig.\ \ref{fig5} by the sequence of snapshots. We show the $z$-component of the magnetization. During its motion, it snakes around an impurity after being pressed against it and thereby being compressed. At a later time, the already compressed Skyrmion gets compressed further, before it eventually disappears. 

\begin{figure}[t!]
	\includegraphics[width=0.99\linewidth]{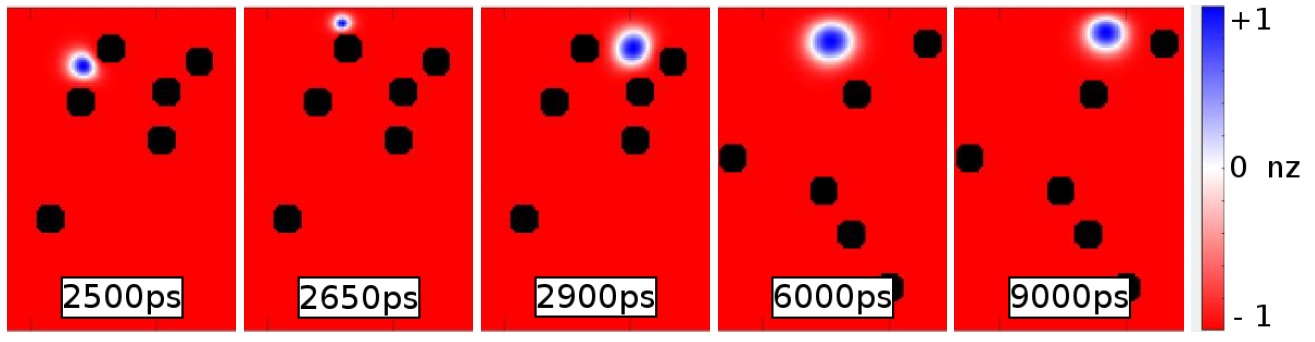}
\caption{\label{fig6} Same as Fig.\ \ref{fig5}, but for $\beta/\alpha = 0.5$,  $v_{\text{s}} = -0.05$, $C = 0.05$ and the radius $R = 6$.  At $t = 2500\text{ps}$,  $t = 2650\text{ps}$ and $t = 2900\text{ps}$, the Skyrmion approaches the upper border due to the Skyrmion Hall effect. Another impurity blocks its further movement such that it gets pinned. }
\end{figure}

Another case of a Skyrmion move with eventual pinning is shown in Fig.\ \ref{fig6}, which typically occurs at a higher impurity concentration. In addition, this case shows a transversal motion due to the Skyrmion Hall effect ($\beta/\alpha = 0.5$). It approaches the upper boundary at $t = 2500\text{ps}$,  $t = 2650\text{ps}$ and $t = 2900\text{ps}$ and another impurity blocks the horizontal motion, such that it permanently stays at its place. 

\begin{figure*}[t!]
	\includegraphics[width=0.7\linewidth]{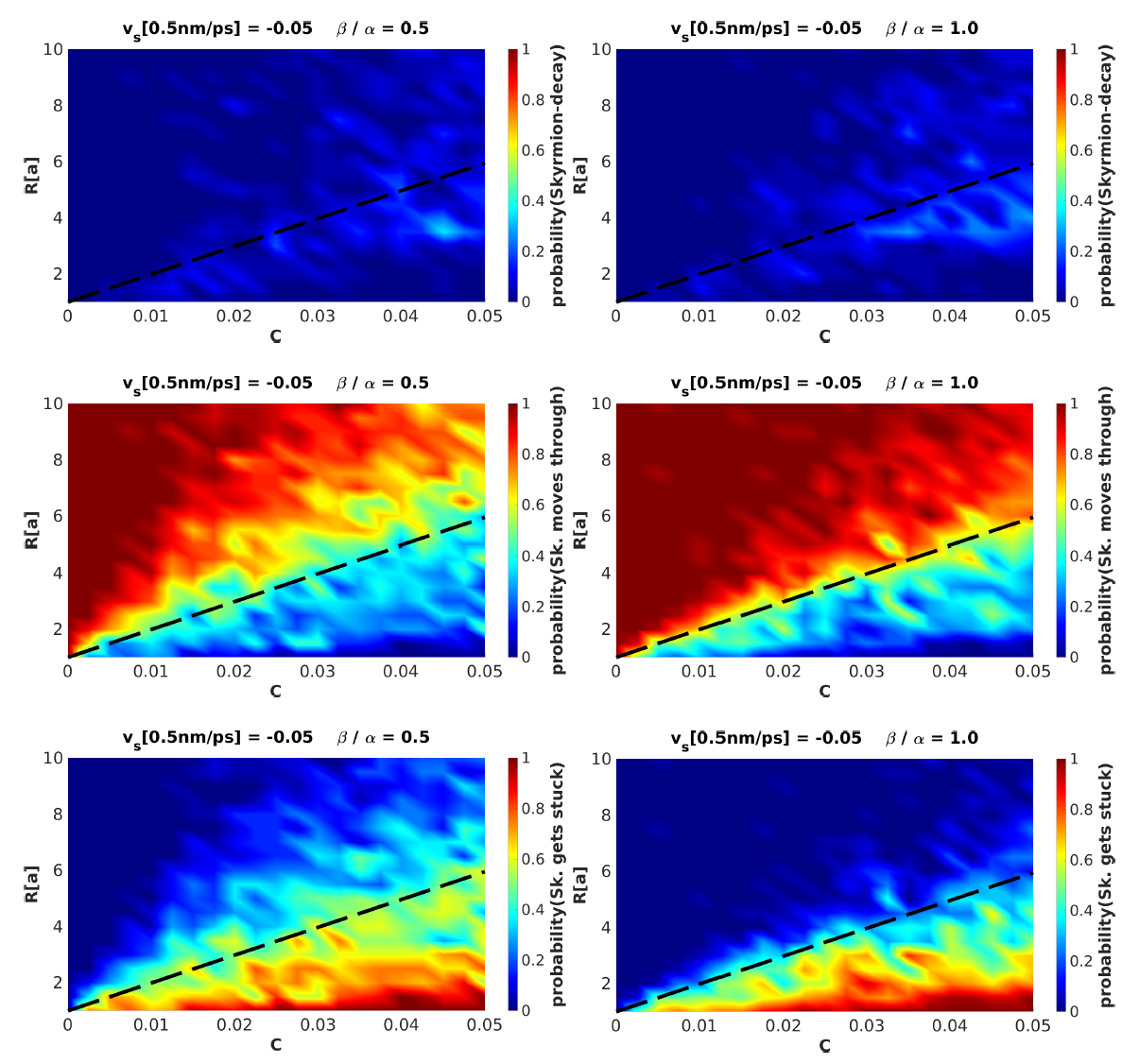}
\caption{\label{fig7} Relative frequency of different types of Skyrmion motion for $\alpha = 0.1$ and $\beta$ as indicated (left column with Skyrmion Hall effect, right column without), obtained for 25 impurity configurations for each data point. Note that the data points below the black dashed lines correspond to regions in parameter space where the charge conservation is significantly violated ($>10 \%$), see discussion in Sec.\ \ref{inhomo}.}
\end{figure*}

A systematic study for varying impurity radii and concentrations yields to the results shown in Fig.\ \ref{fig7}. For this, we have calculated the Skyrmion dynamics for 25 different disorder configurations and have determined the relative frequency of the occurrence of the different types of behavior. In absence of the Skyrmion Hall effect ($\beta=\alpha$, right column), we find that the probability of a Skyrmion decay increases for increasing concentration. Likewise, the probability of the Skyrmion passing is reduced, but is still large. A Skyrmion pinning is not very likely. Almost all find their way around obstacles. Creation of Skyrmions is very unlikely (not shown) and is not considered further here. The presence of the Skyrmion Hall effect ($\beta=0.5 \alpha$, left column) increases the probability of Skyrmion pinning and decreases that of Skyrmion passing. This is due to the additional transverse motion which implies that it will eventually reach a border. 

\begin{figure*}[t!]
	\includegraphics[width=0.7\linewidth]{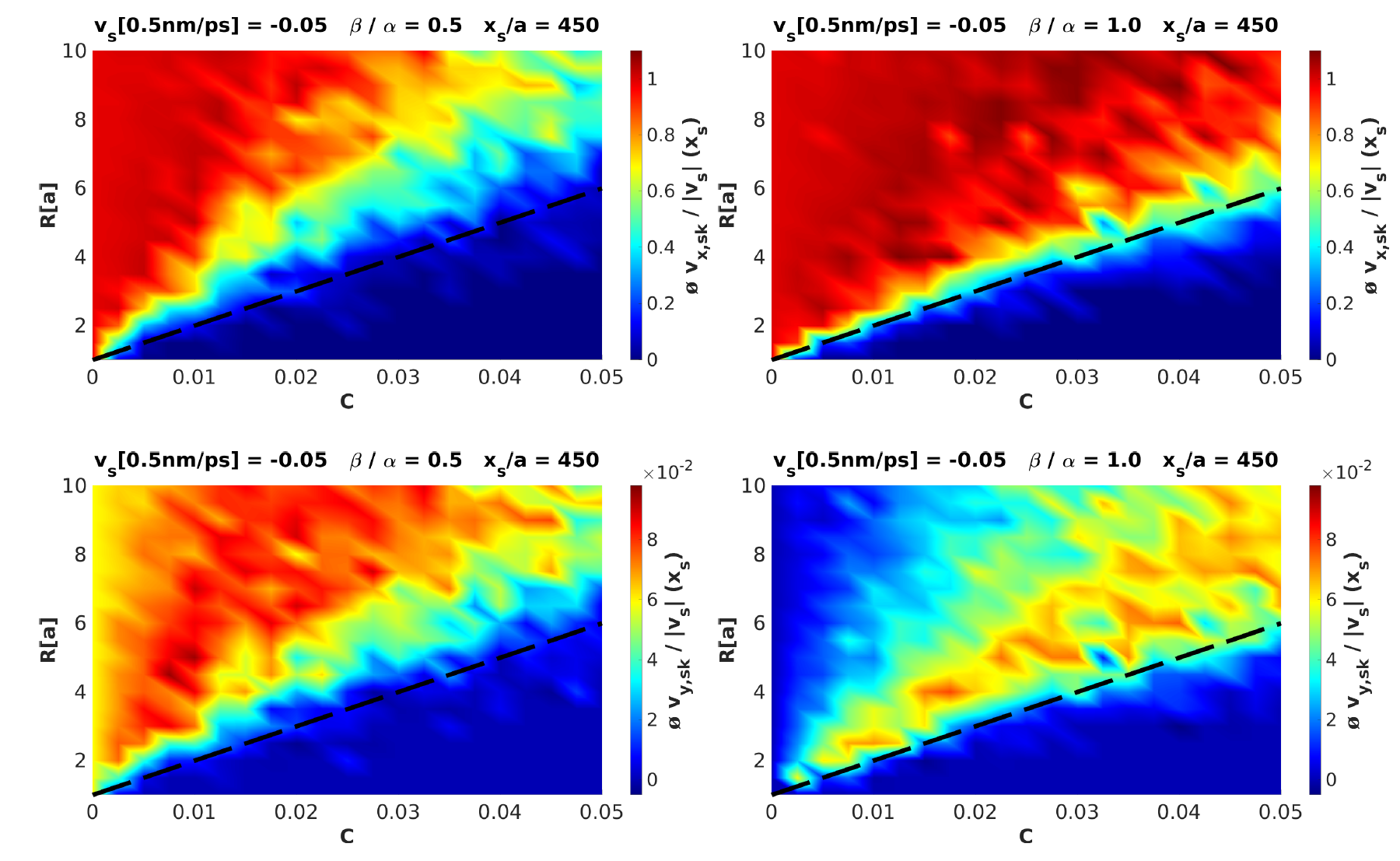}
\caption{\label{fig8} Average values of the normalized velocity components  $v_{x,\text{sk}}$ (top row) and $v_{y,\text{sk}}$ (bottom row) at the longitudinal coordinate $x_{\text{s}}=450 a$ for $\beta=0.5\alpha$ (left column) and  $\beta=\alpha$ (right column) for  $\alpha = 0.1$. For each data point, we have averaged over 25 disorder configurations. The data points below the black dashed lines correspond to regions in parameter space where the charge conservation is significantly violated ($>10 \%$), see discussion in Sec.\ \ref{inhomo}.}
\end{figure*}

Next, we study the Skyrmion velocities characterized by the longitudinal ($v_{x,\text{sk}}$) and transversal ($v_{y,\text{sk}}$) components. To determine these values, we measure the time  $t_{\text{s}}$ at which the Skyrmions reach a specified longitudinal coordinate  $x_{\text{s}}=450 a$.  The results are shown in Fig.\ \ref{fig8}. In general, the transversal component is smaller by two orders of magnitude.  
In the absence of the Skyrmion Hall effect ($\beta=\alpha$, right column), it is naturally very small. A scattering of the Skyrmion by impurities does not alter this significantly, although the longitudinal velocity is reduced. The Skyrmion Hall effect ($\beta=0.5 \alpha$, left column) increases the transversal component slightly, and yields to a stronger deceleration of increasing impurity concentrations. 

\section{Magnetic impurity clusters}

In this section, we consider magnetic impurity clusters. They are characterized by non-zero magnetic anisotropy $K({\bf r})$ in Eq.~(\ref{eq:hamiltonOperator}), their concentration $C$ and the radius $R$. Positive anisotropy, $K>0$, implies a preferential orientation of the magnetization vectors parallel or antiparallel to the $z$-axis. The magnetic impurities are assumed to be conducting, such that the current density is homogeneous over the entire racetrack stripe. Hence, no additional calculation of the current density distribution has to be made and thus no additional discretization errors occur as it was the case for the non-conducting impurities. It is fixed to $v_{\text{s}} = -0.05$, corresponding to $j_c = 6.4 \times 10^{10}\frac{\text{A}}{\text{m}^{2}}$. As for the nonconducting impurities, the mean distance between the clusters increase by reducing $C$ or by increasing $R$. We investigate radii between small ($R = 2$) and large ($R = 5$) values. 

\begin{figure*}[t!]
	\includegraphics[width=0.7\linewidth]{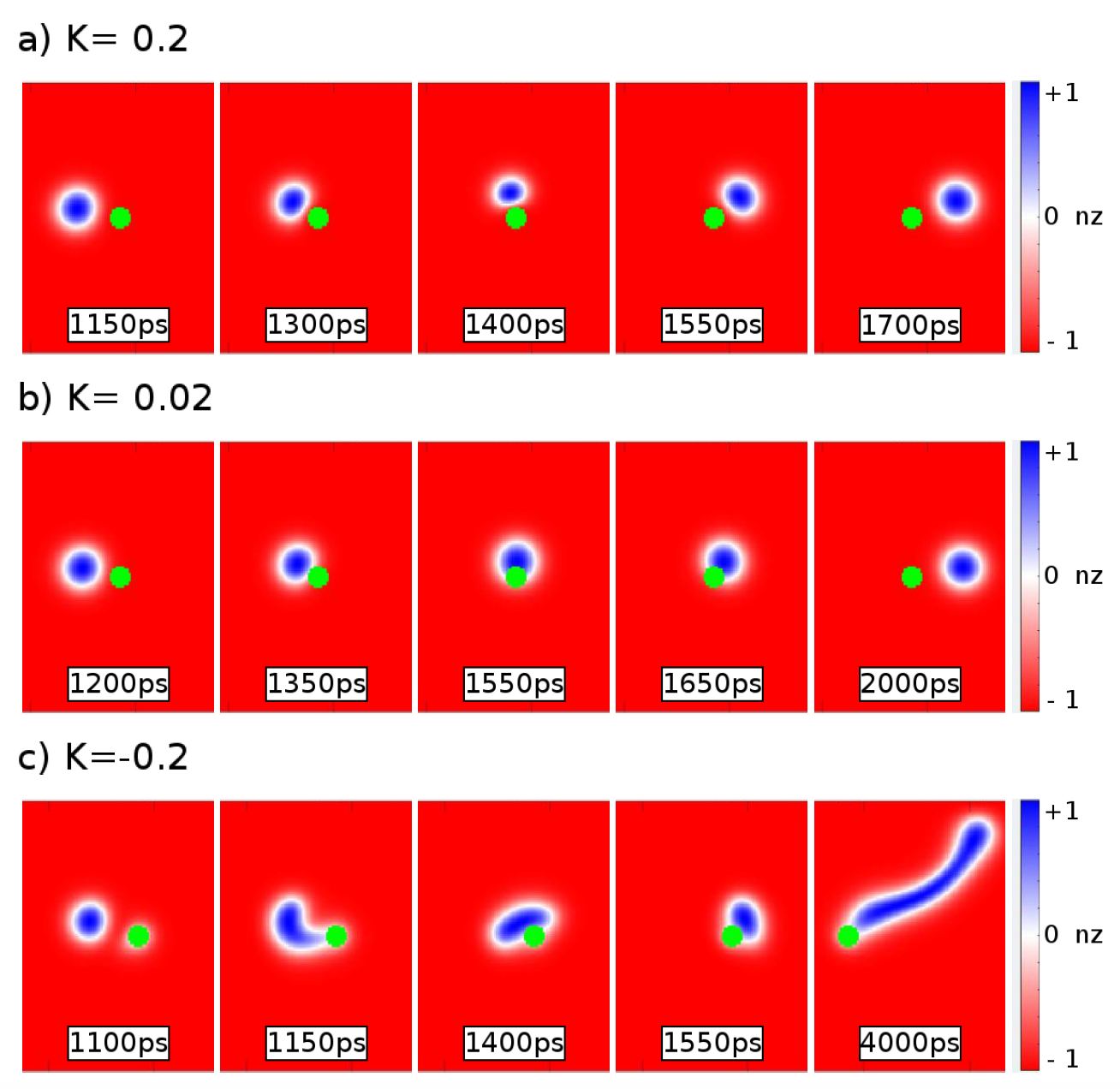}
\caption{\label{fig9} Snapshots of the $z$-component of the magnetization for racetracks with different values of the magnetic disorder anisotropy constant $K$ as indicated for $\beta=\alpha$ (i.e., no Skyrmion Hall effect) and $v_{\text{s}} = -0.05$. The ferromagnetic host is shown in red, the impurity clusters with $R=5$ in green and the moving Skyrmion in blue. }
\end{figure*}

For very large values of $K$, it is energetically difficult to tilt the moments away from the preferential direction, such that impurity clusters appear as rigid objects for the moving textures. A Skyrmion can then move around impurity clusters as shown in Fig.\ \ref{fig9}a. By contrast, for small values of $K$, the Skyrmion can rather easily penetrate the impurity, see Fig.\ \ref{fig9}b for an example. For intermediate values, a mixture of behaviors is encountered and will be analyzed further below. Impurities with a negative anisotropy $K<0$ lead to an easy plane preferential orientation of the magnetic moments in arbitrary directions in the $xy$-plane inside the cluster regions. For very large negative values, only the edge region of the Skyrmion, where its magnetic moments have no finite $z$-component, can interact with the impurity to alter its magnetization. Such an example is shown in 
Fig.\ \ref{fig9}c. We observe a snake-like deformation, an elongation of the Skyrmion and a pinning.  

\subsection{Skyrmion creation}

\begin{figure*}[t!]
	\includegraphics[width=0.7\linewidth]{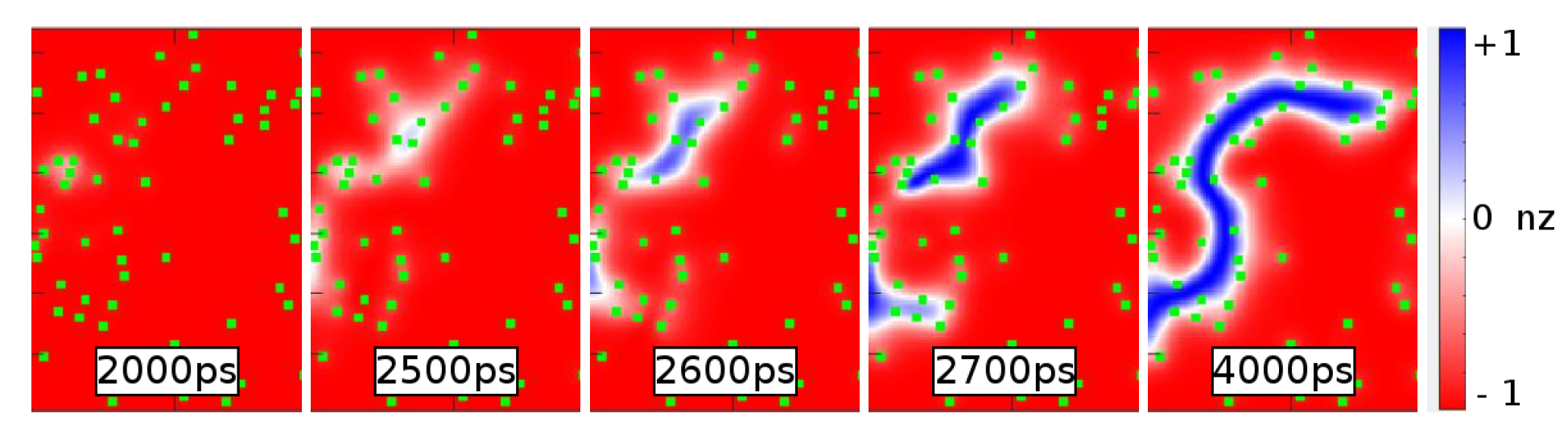}
\caption{\label{fig10} Snap shots of the $z$-components of the magnetization with magnetic impurity clusters with $K=-0.2$ for $C = 0.06$ and $R=2$. Here, no initial Skyrmion was inserted and no current is applied. New Skyrmions spontaneously form at $t = 2000 \text{ps}$. Between neighboring impurity clusters, connections appear (white lines, $t = 2500 \text{ps}$). In the resulting interior region, the magnetization suddenly switches from 
$-{\bf e}_{z}$ to  $+{\bf e}_{z}$ and full Skyrmions appear ($t = 2600 \text{ps}$ and $t = 2700 \text{ps}$). New Skyrmions often have an elongated shape which complies with the disorder configuration ($t = 4000 \text{ps}$).}
\end{figure*}

\begin{figure*}[t!]
	\includegraphics[width=0.7\linewidth]{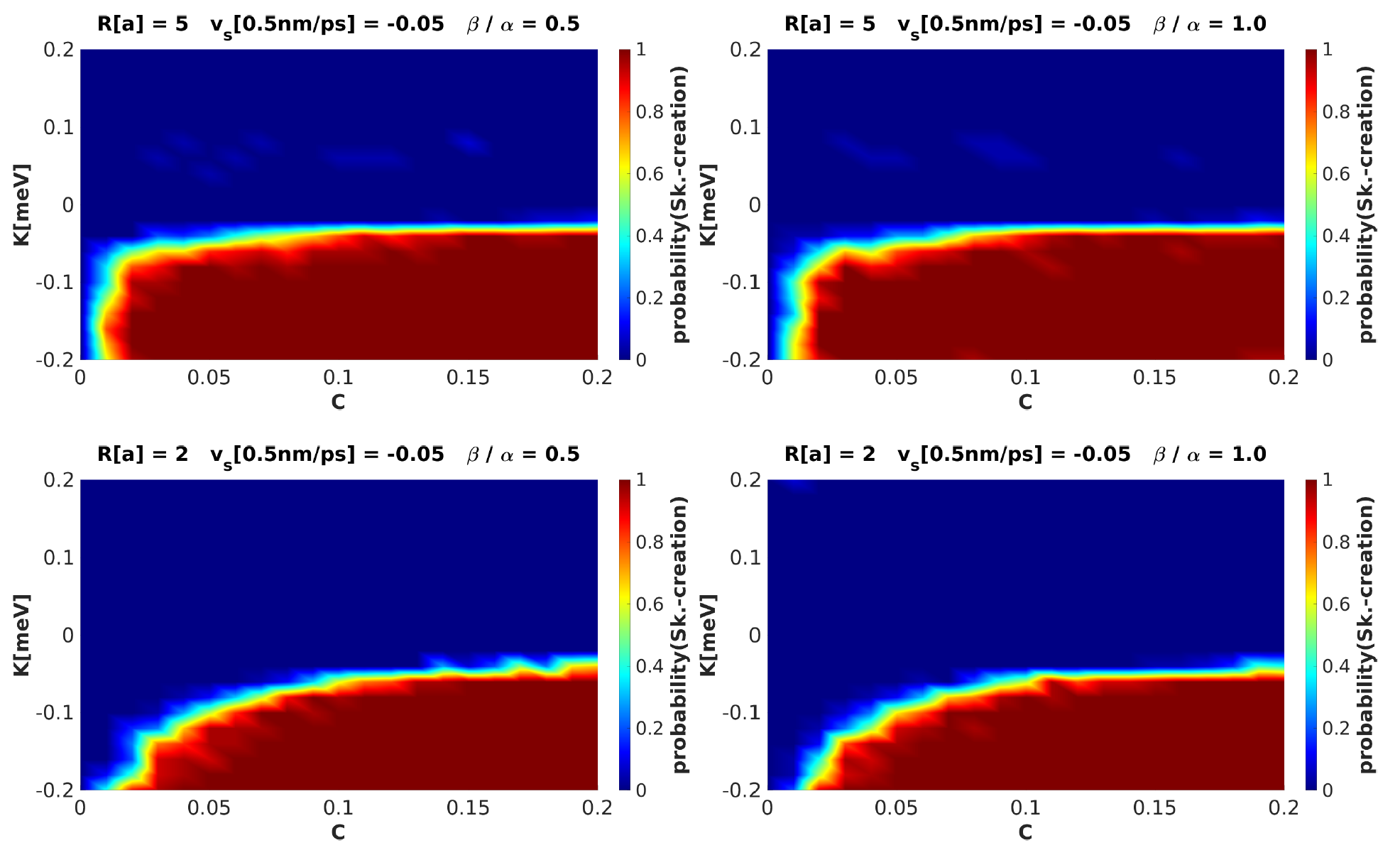}
\caption{\label{fig11}Relative frequency of the creation of new Skyrmions for different values of $R$ and $\beta$ as indicated. We have fixed $\alpha = 0.1$ and have averaged over 25 disorder configurations for each data point.  }
\end{figure*}

For magnetic disorder with negative anisotropy constant $K<0$, we observe that, with very large probability, new Skyrmions are created in the racetrack stripe. An example of a related time sequence is shown in Fig.\ \ref{fig10}. We find certain threshold values of negative anisotropy $K$, concentration $C$ and impurity radius $R$. Then, the probability of Skyrmion creation is close to 1, see 
Fig.\ \ref{fig11}.

Skyrmions can be created also spontaneously and without the presence of an initial seed Skyrmion and/or an applied current as shown in Fig.\ \ref{fig10}. For their creation, a sufficiently large concentration of magnetic disorder with sufficiently large negative anisotropy constant is sufficient. Then, local regions of planar $xy$ magnetizations form spontaneously, which connect between neighboring impurity clusters and which form precursors of Skyrmion edges. In the interior of these regions, the magnetization suddenly switches from 
$-{\bf e}_{z}$ to  $+{\bf e}_{z}$ and full Skyrmions are created. This is independent of the presence of the Skyrmion Hall effect ($\beta =0.5 \alpha$ vs. $\beta = \alpha$ in Fig.\ \ref{fig11}). A large impurity radius, closer to the radius of $R_{\text{sk}} \approx 7.7a$ of the static Skyrmion at equilibrium, is beneficial for the creation also at smaller concentrations and anisotropies. But then, a seed Skyrmion and/or an applied current is required. 

This result is consistent with previous works \cite{stier,everschor}.  New Skyrmions are created in the presence of weak spatial modulations of an external magnetic field \cite{stier}, which generate local magnetization gradients. They facilitate the formation of Skyrmion-Antiskyrmion pairs with zero net topological charge. The Antiskyrmion is energetically unstable and decays, such that only the Skyrmion partner survives. In a related work \cite{everschor}, it was shown that a single magnetic impurity with a modified anisotropy can induce the formation of complicated magnetic textures when a strong spin-polarized current is applied. The current separates the regions of different topological charge by the Skyrmion Hall effect and moves them away from the impurity. Also, the Antiskyrmion decays. The DMI is only required to stabilize the resulting Skyrmion, it is not necessary for its formation. 

\subsection{Skyrmion decay}

\begin{figure*}[t!]
	\includegraphics[width=0.7\linewidth]{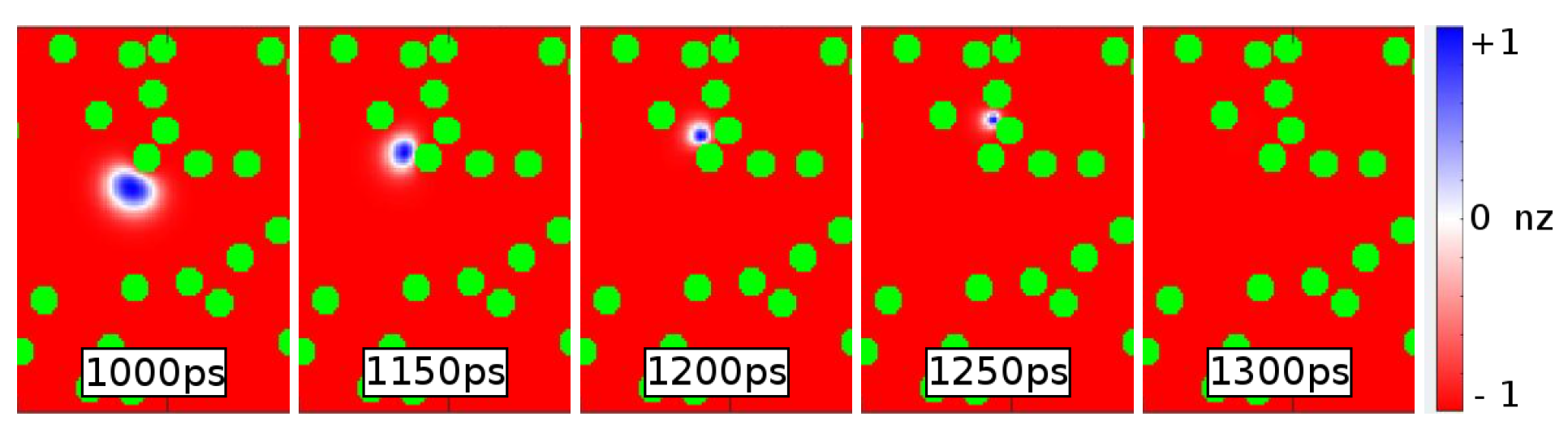}
\caption{\label{fig12}Snapshots of the magnetic $z$-component of a Skyrmion in a racetrack for $\beta=\alpha$, $v_{\text{s}} = -0.05$,  $C = 0.16$,  $R = 5$ and $K=0.2$. The Skyrmion encircles an impurity, gets compressed against others and, eventually, reduces its radius until it disappears. }
\end{figure*}

\begin{figure*}[t!]
	\includegraphics[width=0.7\linewidth]{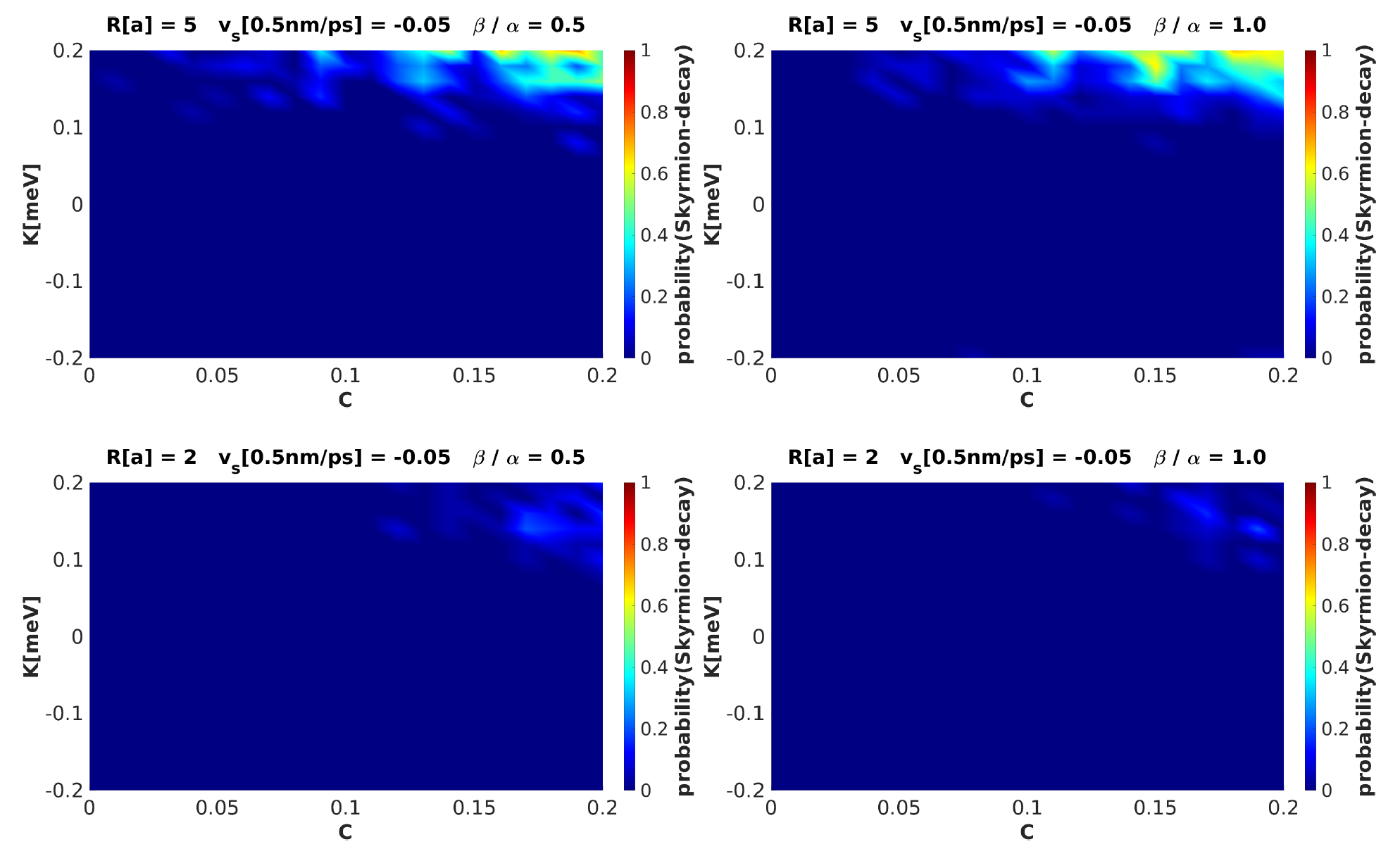}
\caption{\label{fig13}Relative frequency of a Skyrmion decay for different values of the radius and $\beta/\alpha$ as indicated with $\alpha = 0.1$. }
\end{figure*}

When the magnetic disorder is large with a large positive value of the anisotropy constant, the impurity clusters behave like rigid obstacles whose magnetic $z$-component is almost not altered due to the large energy barrier involved. Then, a Skyrmion may be pressed against an impurity, gets compressed and reduced in size, before it eventually decays. Such a case is shown in Fig.\ \ref{fig12}. Such a behavior is more likely for larger concentrations, but requires the special case that impurity clusters lie close to each other. This is yet quite rare and a systematic study shows that the probability of a complete Skyrmion decay is much smaller than 1, see Fig.\ \ref{fig13}. The Skyrmion Hall effect has no particular impact on the relative frequency of such a decay. Large values of the radius typically yield much more frequently to a Skyrmion decay, the reason being that large impurity clusters are more rigid and require more energy to alter their magnetization. 

\subsection{Skyrmion pinning}

For positive anisotropy, $K>0$, and sufficiently large impurity concentration $C$, Skyrmions may get pinned in the racetrack. We find two typical pinning scenarios. 

\begin{figure*}[t!]
	\includegraphics[width=0.7\linewidth]{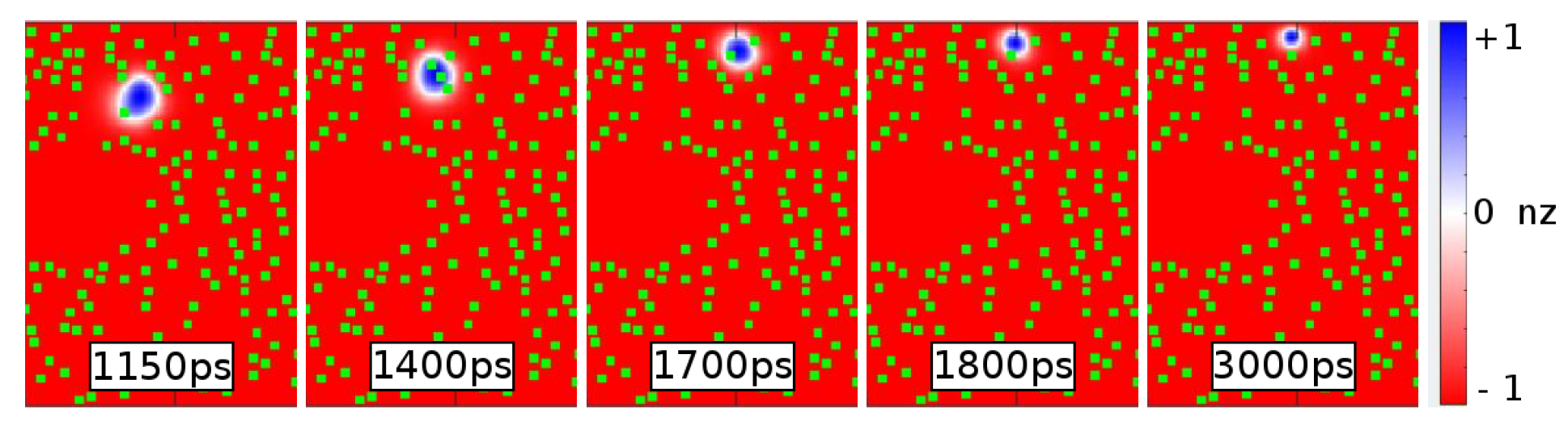}
\caption{\label{fig14}Snapshots of the magnetic $z$-component of a Skyrmion in a racetrack for $\beta=\alpha$, $v_{\text{s}} = -0.05$,  $C = 0.12$,  $R = 2$ and $K=0.16$. The Skyrmion gets caught between several impurity clusters and the lateral edge of the racetrack. }
\end{figure*}

For rather large positive values of $K$, the magnetization of the impurity clusters is relatively rigid. Then, the Skyrmions have to sidestep and might eventually get caught between several impurity clusters and the lateral edge of the racetrack. Such a case is shown in Fig.\ \ref{fig14}.

\begin{figure*}[t!]
	\includegraphics[width=0.7\linewidth]{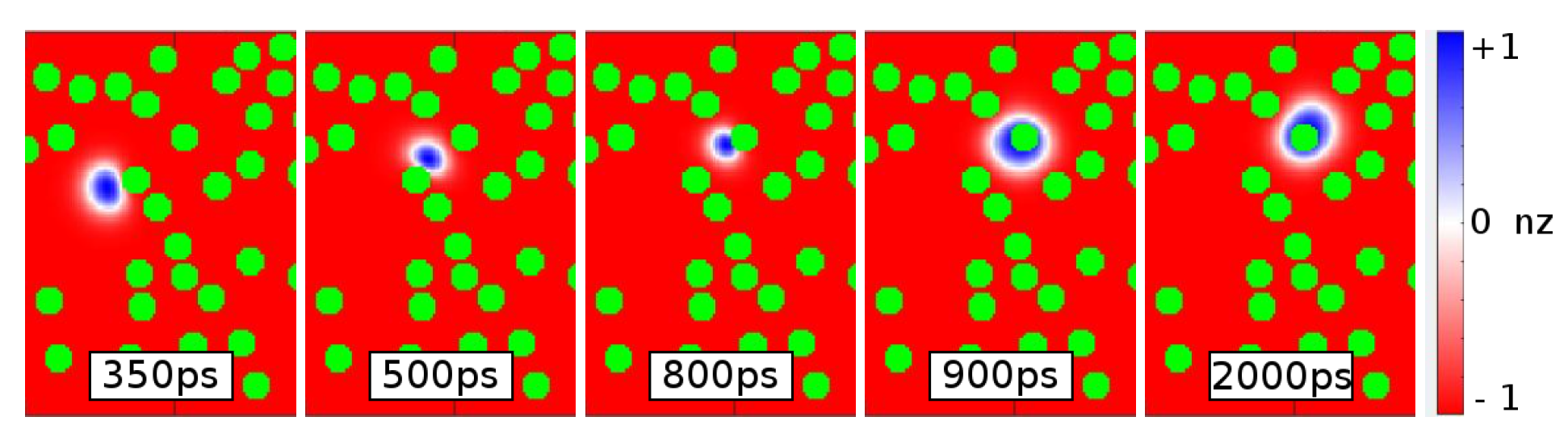}
\caption{\label{fig15}Snapshots of the magnetic $z$-component of a Skyrmion in a racetrack for $\beta=\alpha$, $v_{\text{s}} = -0.05$,  $C = 0.18$,  $R = 5$ and $K=0.12$. The Skyrmion gets caught between several impurity clusters and the lateral edge of the racetrack. The Skyrmion is pressed against a weak magnetic impurity, changes its magnetization from  $+{\bf e}_{z}$ to $-{\bf e}_{z}$ ($t = 900\text{ps}$) and gets pinned at the impurity. }
\end{figure*}

 For smaller positive values of $K$, the less rigid impurity clusters can be passed as their magnetization is dynamically altered. When the change of its magnetization is not perfect and when the Skyrmion center with 
$+{\bf e}_{z}$ magnetization coincides with the impurity, it may get pinned. In this case, both the 
$+{\bf e}_{z}$ and the $-{\bf e}_{z}$ magnetization are energetically equivalent. Such a scenario is shown in Fig.\ \ref{fig15}. In the first scenario, the Skyrmion is caught between several impurity clusters, while in the second, it is directly pinned at the location of one impurity. 

\begin{figure*}[t!]
	\includegraphics[width=0.7\linewidth]{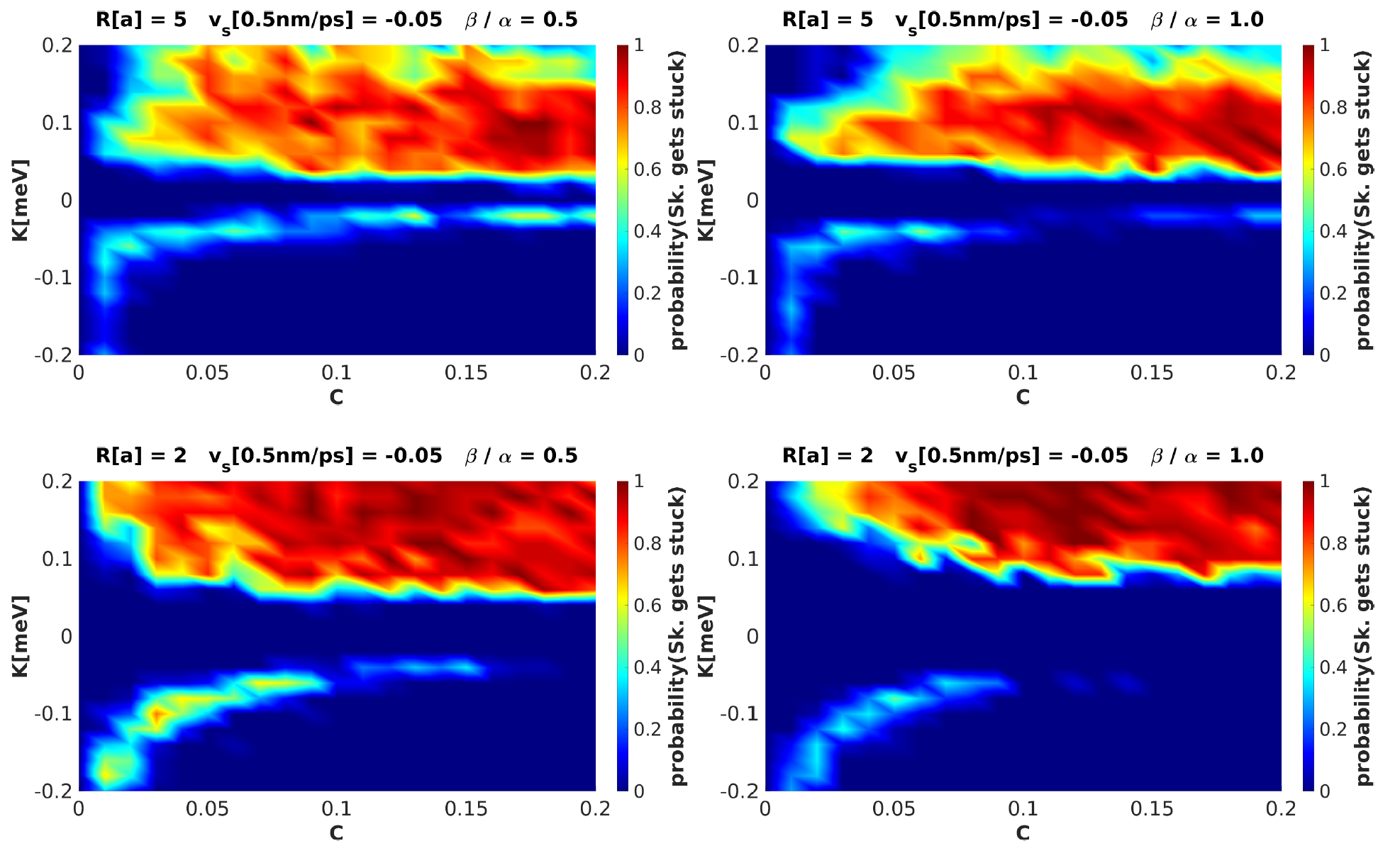}
\caption{\label{fig16}Relative frequency of Skyrmion pinning for various values of $R$ and $\beta$ for $\alpha = 0.1$. }
\end{figure*}

Skyrmion pinning of the second kind is found already for rather low positive values of the anisotropy, but requires a threshold concentration $C$, see Fig.\ \ref{fig16}. The probability of getting pinned increases then with growing $C$. For larger values of $K$ the probability for the first scenario to occur grows. Yet, it usually is below 1, because the decay of the Skyrmion and the passing are competing scenarios. For smaller values of the radius ($R=2$), the tendency of pinning occurs at larger values of $K$, because smaller impurity clusters represent a smaller energy barrier of a change of magnetization than in the case of larger radii  ($R=5$). Once the pinning threshold is reached, the probability of pinning is commonly large for the smaller impurity clusters and for the larger ones. Smaller impurity clusters show a larger effective concentration for the same value of $C$, such that a Skyrmion has by default fewer possibilities to move around. 

In the presence of the Skyrmion Hall effect, the probability of pinning increases, in particular already at lower impurity concentration. In this case, the Skyrmion has a transversal velocity component and the chance to detour impurity clusters decreases relative to the case without the Skyrmion Hall effect. 

It may also occur for negative values of $K$ that Skyrmions get pinned. In this case, the magnetization vectors of the impurity clusters prefer directions with the $xy$-plane. Such orientations occur in the interior regions of a Skyrmion, which connect the center and the edge. Then, a Skyrmion can be slowed down or get stuck by pinning to such impurity clusters with these regions.  

\subsection{Skyrmion passing}

\begin{figure*}[t!]
	\includegraphics[width=0.7\linewidth]{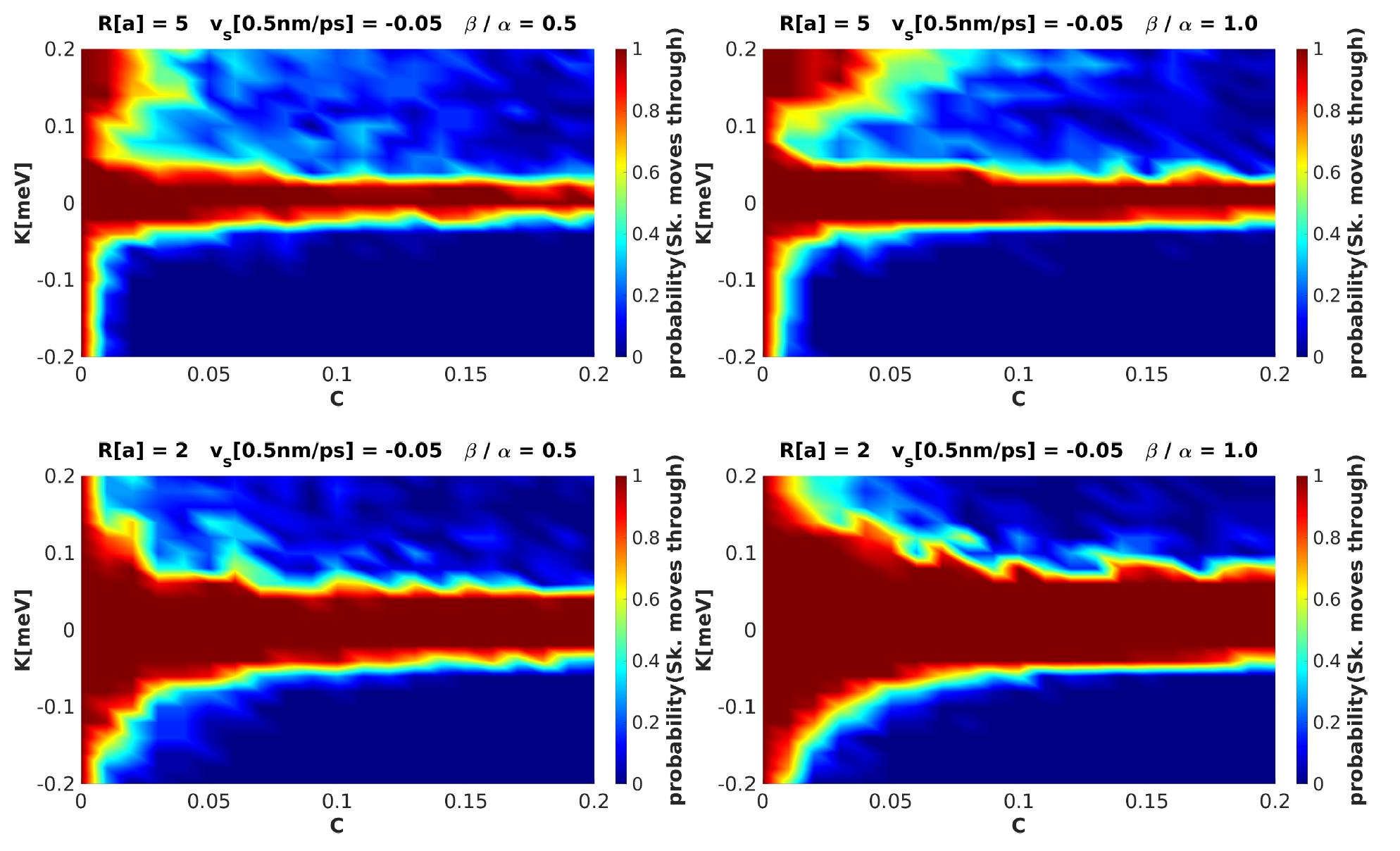}
\caption{\label{fig17}Relative frequency of the passing of Skyrmions for $\alpha = 0.1$. }
\end{figure*}

Ordinary passage of the racetrack by the Skyrmion is the most common behavior if the perturbations discussed so far do not change its motion. This type of behavior dominates for low impurity concentration $C$ and low anisotropy constants, see Fig.\ \ref{fig17}. Disorder then is too weak to alter the global longitudinal motion. For negative anisotropy, the creation of new Skyrmions and pinning limits the longitudinal passing. Also for positive values of $K$, pinning may occur, but is then less pronounced. Then, longitudinal motion is possible in general for all positive values of $K$, albeit with reduced probability.

For smaller impurity clusters $R=2$, the probability of Skyrmion transmission is larger. It is in general large up to a certain threshold value of the anisotropy and decreases then rapidly, because the other types of behaviors then prevail. The Skyrmion Hall effect has an impact inasmuch it reduces the probability of complete transmission. The enforced transversal motion lowers the number of possibilities to move around the impurity clusters. Hence, the probability of pinning grows. 

\subsection{Summary and synopsis}

\begin{figure*}[t!]
	\includegraphics[width=0.7\linewidth]{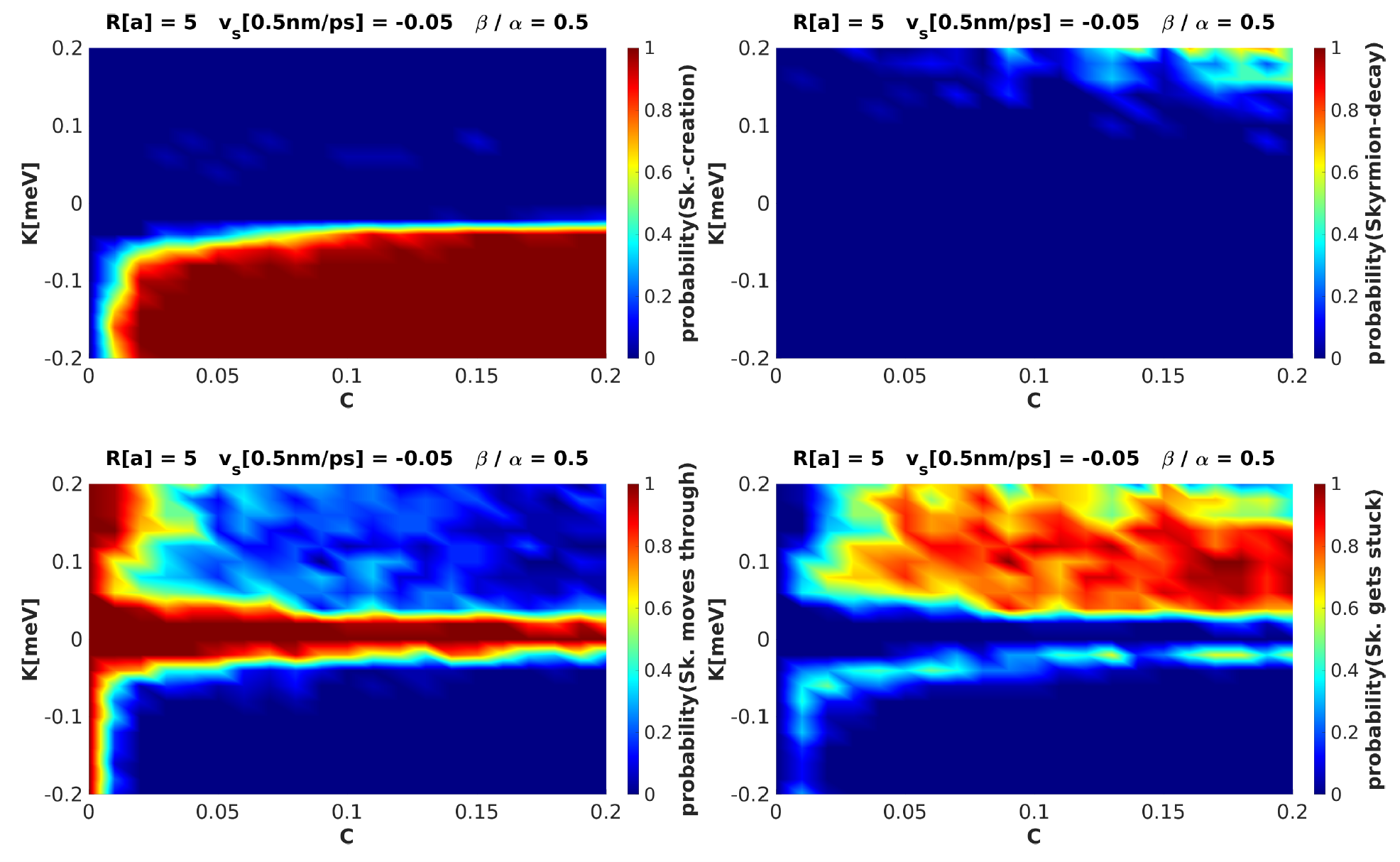}
\caption{\label{fig18}Synopsis of the relative frequencies of the four types of Skyrmion behavior for $R=5$ and $\beta= 0.5\alpha$ with $\alpha = 0.1$. }
\end{figure*}

\begin{figure*}[t!]
	\includegraphics[width=0.7\linewidth]{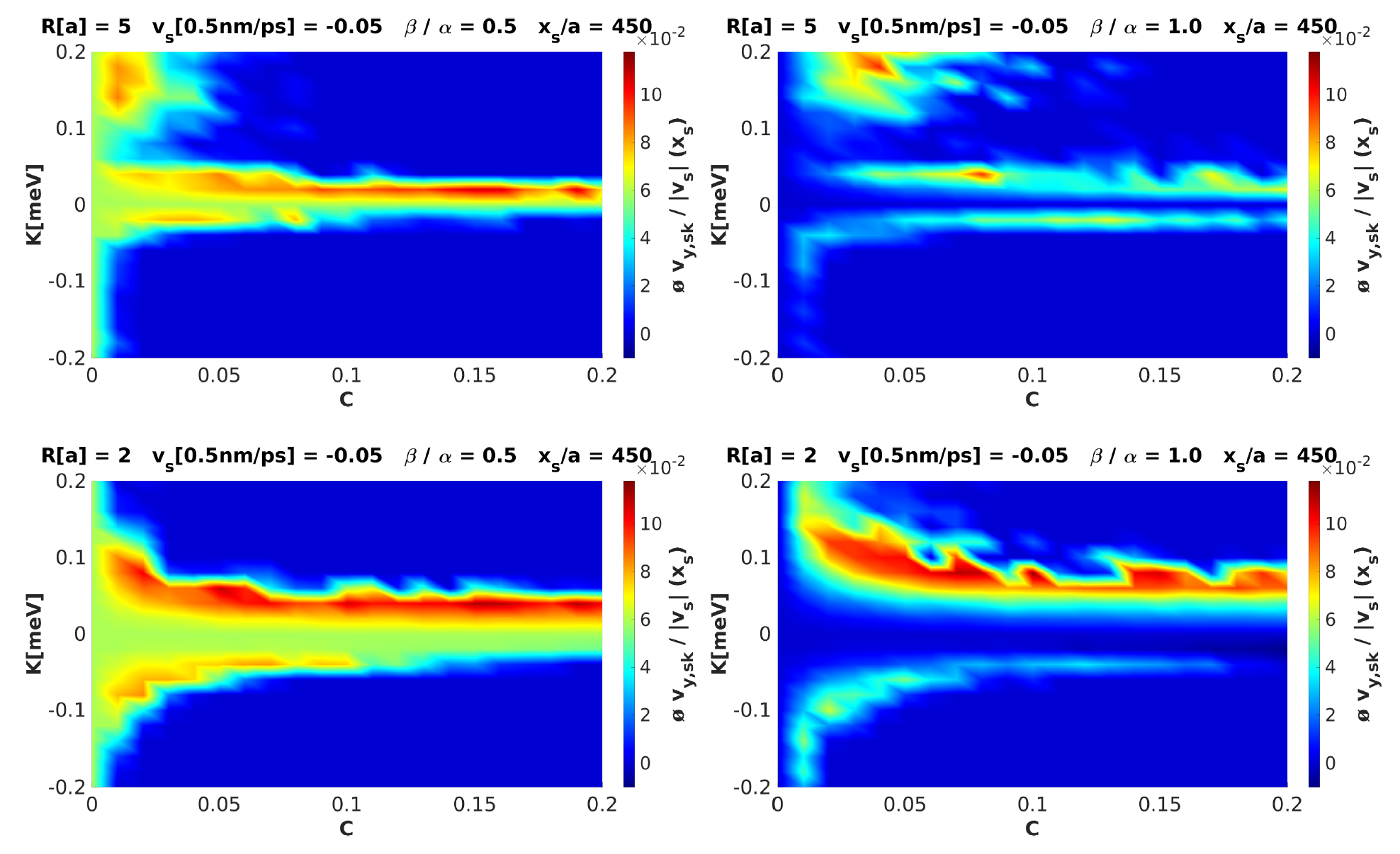}
\caption{\label{fig19}Mean average transverse velocity $v_{y,\text{sk}}$ normalized to the entrance velocity 
 $|v_{\text{s}}|$ which occurs when the coordinate $x_{\text{s}}$ is reached for  $\alpha = 0.1$. The remaining parameters are as indicated.  }
\end{figure*}

A synopsis of the different types of Skyrmion motion is shown in Fig.\ \ref{fig18}. Certainly, a maximal transmission of Skyrmions requires minimal disorder and a minimal value of the anisotropy. A transversal motion with a transversal velocity component $v_{y,\text{sk}}$ can have two reasons, see Fig.\ \ref{fig19}. The first one is the ordinary Skyrmion Hall effect ($\beta / \alpha = 0.5$). The second one occurs also in absence of the Skyrmion Hall effect ($\beta = \alpha$) when many and strong enough impurity clusters are present. Then, Skyrmions can be forced to pass some transversal paths due to scattering at the impurity clusters and lateral passing. In general, the positive $y$ direction is preferred against the negative one. This is similar to the observations of the non-conducting impurity clusters and might be due to the symmetry breaking induced by the choice of Bloch Skyrmions with a given helicity $\gamma = -\pi / 2$ or $\gamma = +\pi / 2$. 

\section{Conclusions}
\label{sec:Conclusions}

We have studied the current-driven dynamics of Skyrmions in a ferromagnetic racetrack with disordered impurity clusters. The latter are assumed either non-conducting or magnetic. We have solved the generalized Landau-Lifshitz-Gilbert equation in the presence of a current-induced spin transfer torque. For non-conducting impurity clusters, the inhomogeneous current density distribution has been calculated by solving the basic electrostatic equations. For the magnetic (conducting) impurity clusters, the applied homogeneous current density is present all over the racetrack stripe. In dependence of the disorder concentration and the radius of the impurity clusters, three different characteristic behaviors result in addition to the standard desired Skyrmion passing. Induced by the impurity clusters as well as by the current, new Skyrmions can be created in the stripe. Moreover, Skyrmions can get pinned at the impurity clusters, and Skyrmions can decay after being pressed against individual impurity clusters.  

For growing concentration of non-conducting impurities, the probabilities of pinning and decay grow. The same is true for magnetic impurities and a growing anisotropy constant.  Depending on the sign of the anisotropy, the probabilities of pinning and decay grows (preferably for $K>0$), or new Skyrmions are created (for $K<0$). 

Our results demonstrate that a careful tuning of the parameters permits very different dynamical behavior of Skyrmions in disordered racetracks. \\

\section*{Acknowledgments} We gratefully acknowledge funding by the Deutsche Forschungsgemeinschaft (project number 403505707) within the DFG SPP 2137 ``Skyrmionics''.


\begin{thebibliography}{99}

\bibitem{Bogdanov1995}
A. Bogdanov, New localized solutions of the nonlinear field equations, Sov. Phys. JETP Lett. \textbf{62}, 247 (1995).

\bibitem{bogdanov1}
A.N. Bogdanov and D.A. Yablonski\u{\i}, Thermodynamically stable vortices in magnetically ordered crystals. The mixed state of magnets, Sov. Phys. JETP \textbf{68}, 101 (1989).

\bibitem{pfleidererrosch2}
S. Mühlbauer, B. Binz,  F. Jonietz,  C. Pfleiderer,  A. Rosch, A. Neubauer, R. Georgii, and P. Böni,  Skyrmion Lattice in a Chiral Magnet, Science \textbf{323}, 915 (2009).

\bibitem{yu1}
X. Yu, N. Kanazawa, Y. Onose, K. Kimoto, W. Zhang, S. Ishiwata, Y. Matsui, and Y. Tokura, Near room-temperature formation of a skyrmion crystal in thin-films of the helimagnet FeGe, Nat. Mater. \textbf{10}, 106 (2011).

\bibitem{heinze}
S. Heinze, K. von Bergmann, M. Menzel, J. Brede, A. Kubetzka, R. Wiesendanger, G. Bihlmayer, and S. Blügel,  Spontaneous atomic-scale magnetic skyrmion lattice in two dimensions, Nat. Phys. \textbf{7}, 713 (2011).

\bibitem{nagaosa1}
J. Iwasaki, M. Mochizuki, and N. Nagaosa,  Universal current-velocity relation of skyrmion
motion in chiral magnets, Nat. Commun. \textbf{4}, 1463 (2013).

 \bibitem{Lu2014}W. Lu and J. Xiang, {\textit Semiconductor Nanowires: From Next-Generation Electronics to Sustainable Energy} (The Royal Society of Chemistry, Cambridge, 2015). 

\bibitem{romming}
N. Romming, C. Hanneken, M. Menzel, J. Bickel, B. Wolter, K. von Bergmann, A. Kubetzka, and R. Wiesendanger, Writing and deleting single magnetic skyrmions, Science \textbf{341}, 636 (2013).

\bibitem{garanin18}
A. Derras-Chouk, E. M. Chudnovsky, and D. A. Garanin, Quantum collapse of a magnetic skyrmion, Phys. Rev. B \textbf{98}, 024423 (2018).

\bibitem{rosch19}
B. Heil, A. Rosch, and J. Masell,  Universality of annihilation barriers of large magnetic skyrmions in chiral and frustrated magnets, Phys. Rev. B \textbf{100}, 134424 (2019).

\bibitem{fuchsbacharb}
A. Fuchs,  Energy Barrier of a Decaying Magnetic Skyrmion on the Square Lattice. Universität Augsburg, Bachelor Thesis (2020).

\bibitem{stier}
M. Stier, W. Häusler, T. Posske, G. Gurski, and M. Thorwart, Skyrmion-antiskyrmion pair creation by in-plane currents., Phys. Rev. Lett. \textbf{118}, 267203 (2017).

\bibitem{Litzius2017} K. Litzius, I. Lemesh, B. Krüger, P. Bassirian, L. Caretta, K. Richter, F. Büttner, K. Sato, O.A. Tretiakov,J. Förster, R. M. Reeve, Robert, M. Weigand, I. Bykova, H.  Stoll, G. Schütz, G. S. D. Beach, and M. Kläui, Skyrmion Hall effect revealed by direct time-resolved X-ray microscopy, Nat. Phys. {\bf 13}, 170 (2017). 

\bibitem{jiangshe}
W. Jiang, X. Zhang, G. Yu, W. Zhang, X. Wang, M. Jungfleisch, J. Pearson, X. Cheng, O. Heinonen, K. Wang, Y. Zhou, A. Hoffmann, and S. te Velthuis,  Direct observation of the skyrmion Hall effect, Nat. Phys. \textbf{13}, 162 (2017).

\bibitem{realhrabec}
A. Hrabec, J. Sampaio, M. Belmeguenai, I. Gross, R. Weil, S. Cherif, A. Stashkevich, V. Jacques, A. Thiaville,  and S. Rohart,  Current-induced skyrmion generation and dynamics in symmetric bilayers, Nat. Commun. \textbf{8}, 15765 (2017).

\bibitem{realjuge}
R. Juge, S. Je, D. de Souza Chaves, L. Buda-Prejbeanu, J. Pena-Garcia, J. Nath, I. Miron, K. Rana, L. Aballe, M. Foerster, F. Genuzio, T. Mentes, A. Locatelli, F. Maccherozzi, S. Dhesi, M. Belmeguenai, Y. Roussigne, S. Auffret, S. Pizzini, G. Gaudin, J. Vogel, and O. Boulle, Current-Driven Skyrmion Dynamics and Drive-Dependent Skyrmion Hall Effect in an Ultrathin Film,  Phys. Rev. Applied \textbf{12}, 044007 (2019).

\bibitem{realwoo}
S. Woo, K. Litzius, B. Krüger, M. Im, L. Caretta, K. Richter, M. Mann, A. Krone, R. Reeve, M. Weigand, P. Agrawal, I. Lemesh, M. Mawass, P. Fischer, M. Kläui, and G. Beach,  Observation of room-temperature magnetic skyrmions and their current-driven dynamics in ultrathin metallic ferromagnets, Nat. Mater. \textbf{15}, 501 (2016).

\bibitem{barati}
E. Barati, M. Cinal, D. Edwards, and A. Umerski, Gilbert damping in magnetic layered systems, Phys. Rev. B \textbf{90}, 014420 (2014).

\bibitem{sekiguchi}
K. Sekiguchi, K. Yamada, S. Seo, K. Lee, D. Chiba, K. Kobayashi, and T. Ono,  Time-Domain Measurement of Current-Induced Spin Wave Dynamics, Phys. Rev. Lett. \textbf{108}, 017203 (2012).

\bibitem{roessler}
S. Rößler, S. Hankemeier, B. Krüger, F. Balhorn, R. Frömter, and H. Oepen,  Nonadiabatic spin-transfer torque of magnetic vortex structures in a permalloy square, Phys. Rev. B \textbf{89}, 174426 (2014).

\bibitem{everschor}
K. Everschor-Sitte, M. Sitte, T. Valet, A. Abanov, and J. Sinova,  Skyrmion production on demand by homogeneous DC currents, New J. Phys. \textbf{19}, 092001 (2017).



\end{thebibliography}
\end{document}